\documentclass[aps,prc,twocolumn,showpacs,superscriptaddress,footinbib]{revtex4-1}

\usepackage{graphicx}
\usepackage{dcolumn}
\usepackage{bm}
\usepackage{enumerate}
\usepackage{amsmath}

\usepackage[normalem]{ulem}  

\begin{document}


\title{Prolate-oblate asymmetric shape phase transition in the interacting boson model with $SU(3)$ higher-order interactions}

\author{Tao Wang}
\email{suiyueqiaoqiao@163.com}
\affiliation{College of Physics, Tonghua Normal University, Tonghua 134000, People's Republic of China}

\author{Bing-cheng He}
\email{bhe@anl.gov}
\affiliation{Physics Division, Argonne National Laboratory, Lemont, Illinois 60439, USA}

\author{Dong-kang Li}
\email{ldk667788@163.com}
\affiliation{College of Physics, Tonghua Normal University, Tonghua 134000, People's Republic of China}

\author{Chun-xiao Zhou}
\email{zhouchunxiao567@163.com}
\affiliation{College of Mathematics and Physics Science, Hunan University of Arts and Science, Changde 415000, People¡¯s Republic of China}

\date{\today}

\begin{abstract}
\textbf{Abstract}: Prolate-oblate shape phase transition is an interesting topic in nuclear structure, which is useful for understanding the intrinsic interactions between nucleons.  Recently, the interacting boson model with $SU(3)$ higher-order interactions was proposed, in which the prolate shape and the oblate shape are not described in a mirror symmetric way. This asymmetric description seems more realistic. The level evolutions, $B(E2)$ values and other important indicators showing the prolate-oblate asymmetric transitions are investigated in detail, and realistic structure evolutions from $^{180}$Hf to $^{200}$Hg are compared. A key finding is that, the average deformation of the prolate shape is nearly twice the one of the oblate shape. These results, together with the successful description of the $B(E2)$ anomaly in $^{168,170}$Os, $^{172}$Pt, the $\gamma$-soft properties of $^{196}$Pt, $^{82}$Kr and the normal states of $^{110}$Cd, support the validity of the new model.
\end{abstract}

\maketitle

\section{Introduction}

Experimental anomalies have the potential to induce a new understanding of the same problem. In the last decade of researches in the field of nuclear structure, two types of abnormal phenomena have attracted great attentions, both of which are difficult to incorporate into the existing theoretical framework. The most direct deviation from existing research experiences is the $B(E2)$ anomaly phenomenon \cite{168Os,166W,172Pt,170Os,114Xe,114Te,74Zn,50Cr}, in which the ratio of reduced transition probabilities $B_{4/2}=B(E2;4_{1}^{+}\rightarrow 2_{1}^{+})/B(E2;2_{1}^{+}\rightarrow 0_{1}^{+})$ in the yrast band can be much smaller than 1 (a possible non-collective signal) while the energy ratio of the corresponding levels $E_{4/2}=E_{4_{1}^{+}}/E_{2_{1}^{+}}$ is  equivalent to or larger than 2, which is a typical indicator for the emergence of collectivity in nuclear structure. Such a sudden emergence of empirical opposition rejects the explanations of existing theories, such as the interacting boson model (IBM-2) calculations based on the SkM$^{*}$ energy-density functional \cite{168Os} and the symmetry-conserving configuration mixing (SCCM) calculations \cite{170Os}. Another and more important anomaly is the spherical nucleus puzzle (Cd puzzle for the Cd isotopes) \cite{Garrett08,Garrett10,Heyde11,Garrett12,Batchelder12,Heyde16,Garrett18,Garrett19,Garrett20}, in which the vibrational mode of a rigid spherical nucleus is questioned and refuted, and it was suggested that the overall performance of the normal states of the Cd isotopes is a manifestation of one specific $\gamma$-soft rotational mode \cite{Garrett10,Garrett12}. These experimental results are in conflict with our long-held conception on the collective vibrational pattern of spherical nuclei. It means that the nuclei near the closed shells (or even the magic nuclei) may not be spherical \cite{Garrett19,Togashi18}. Understanding the two anomalies is becoming an increasingly important topic in the research of nuclear structure, but meaningful progress is rare, because both disagree with existing experiences. Existing theories do not seem to support the emergence of such phenomena.

\begin{figure}[tbh]
\includegraphics[scale=0.27]{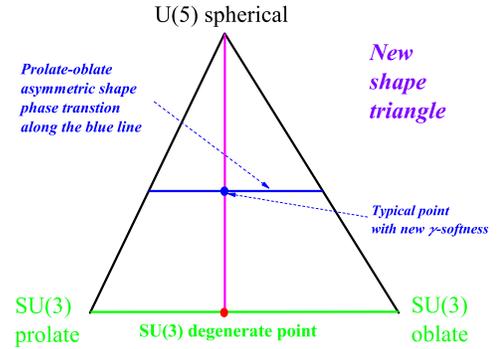}
\caption{The simplest SU3-IBM can be described by the new spherical-prolate-oblate shape triangle. The blue line presents the prolate-oblate shape asymmetric evolutional path discussed in this paper.}
\end{figure}

Recently, the interacting boson model with $SU(3)$ higher-order interactions (SU3-IBM for short) was proposed by one of the authors to try to resolve the two puzzles in a unified way \cite{Wang22,Wang20,Zhang22}. This is a new extension of the previous interacting boson model (IBM) proposed by Arima and Iachello \cite{Iachello87}. In this algebraic model, the $s$ and $d$ bosons are used to construct the Hamiltonian of a nucleus to explain the collective behaviors in low-lying nuclear excitations. These bosons can be regarded as pairs of nucleons with angular momentum $L=0$ and $L=2$. There are three solvable algebraic limits: the $U(5)$ limit presents the spherical vibrational mode, the $SU(3)$ limit presents the prolate shape  (up to two-body interactions) and the $O(6)$ limit presents the $\gamma$-unrelated rotation. There is also a $\overline{SU(3)}$ case presenting the oblate shape (extended Casten triangle \cite{Casten06}). The $SU(3)$ limit and the $\overline{SU(3)}$ case are on either side of the $O(6)$ limit and mirror symmetric about the $\gamma$-unrelated point from the spectra perspective \cite{Jolie03t,Wang08}. In the SU3-IBM, the $SU(3)$ higher-order interactions are exploited, so the $SU(3)$ limit not only describe the prolate shape (second-order Casimir operator), but also presents the oblate shape (third-order Casimir operator) \cite{Wang22}, even the triaxial rigid rotor by introducing the forth-order interactions \cite{Zhang22}. Thus in the new model, oblate shape is described using a new way with $SU(3)$ symmetry, which is different from the one used in the previous IBM with $\overline{SU(3)}$ symmetry. The new spherical-prolate-oblate shape triangle can be seen in Fig. 1.

\begin{figure}[tbh]
\includegraphics[scale=0.27]{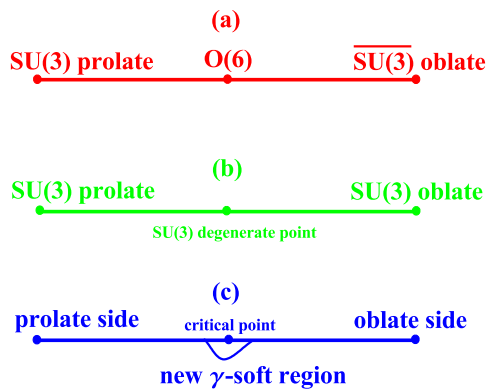}
\caption{Three prolate-oblate transitional paths discussed in the IBM. (a) the $SU(3)$ limit to the $\overline{SU(3)}$ limit via the $O(6)$ symmetry in previous IBM; (b) the $SU(3)$ prolate side to the $SU(3)$ oblate side via the $SU(3)$ degenerate point within the $SU(3)$ limit; (c) the prolate side to the oblate side via the new $\gamma$-soft region with a critical point in the SU3-IBM.}
\end{figure}

This new model based on Fortunato \emph{et al.}'s critical findings \cite{Fortunato11}. They first pointed out that the $SU(3)$ three-body interaction can present the oblate shape. Thus the $\overline{SU(3)}$ description of the oblate shape in previous IBM can be replaced by the $SU(3)$ third-order interaction. However, at the beginning of the result, the substitution did not attract enough attentions. An exception is Zhang \emph{et al.}'s interesting work \cite{Zhang12}. A simple analytic description for prolate-oblate shape phase transition can be given, which is a rare first-order phase transition occurring for finite $N$ (the green line in Fig. 1). The $SU(3)$ limit with higher-order interactions can have various shapes with rigid quadrupole deformation for the ground state of a nucleus \cite{Isacker00,zhang14}, that is, it can describe quadrupole shape coexistence in a simple manner \cite{Heyde11}. This view has not been fully explored yet \cite{communication}. Similar discussions have been performed by Leviatan \emph{et al.} \cite{Leviatan16,Leviatan17}. The shape phase transition point from the $SU(3)$ prolate shape to the $SU(3)$ oblate shape is also a degenerate point \cite{Zhang12}. Fortunato \emph{et al.} pointed out that, in the large $N$ limit, the evolution path from the $U(5)$ limit to the $SU(3)$ degenerate point can have $\gamma$-rigid potential energy surface for the ground state, but the potential along the $\gamma$ degree of freedom is shallow (the magenta line in the Fig. 1) \cite{Fortunato11}. One of the authors (T. Wang) offered a numerical study of this transition line in detail for finite $N$, and one result beyond expectation is that the whole region has a kind of $\gamma$-softness with $O(5)$ partial dynamical symmetry \cite{Wang22}. This result is different from the traditional experience based on the previous IBM and is very fascinating. This $\gamma$-softness is an emergent phenomenon. It is found that, this new $\gamma$-soft rotational mode may be the possible correct description for the normal states of the Cd isotopes. And, we also find that, it can be exploited to explain the $\gamma$-soft feature in $^{196}$Pt \cite{Wang23}, and even the $E(5)$-like $\gamma$-softness in $^{82}$Kr \cite{Zhou23}. This means that we have a new perspective on understanding the properties of realistic $\gamma$-soft nuclei, which has been discussed by many nuclear theories \cite{Jean56,Iachello78,Casten78,Fewell85,Bijker82,Dukelsky04,Caprio04,Caprio05,Nomura12,Isacker10,Bohr75,Ring80,Reinhard03,Zuker05,Ring11,Shimizu12,Nomura21}. In this new theory, the $\gamma$-softness is weakly related to the $\gamma$ geometric variables, but can have exact symmetry partly \cite{Wang22}.

\begin{figure}[tbh]
\includegraphics[scale=0.27]{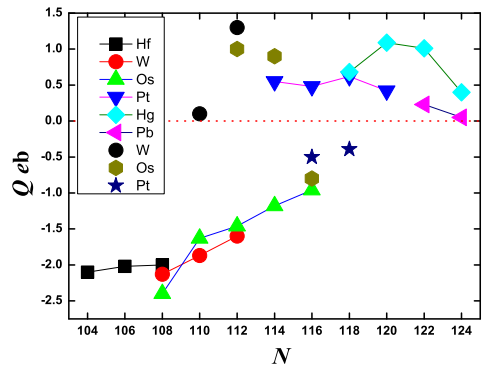}
\caption{The experimental quadrupole moments $Q_{2_{1}^{+}}$ (with connection) and $Q_{2_{2}^{+}}$ (without a connection) of the first and second $2^{+}$ states in the Hf-Pb region.}
\end{figure}

To further confirm the validity of the new theory SU3-IBM and the new emergent $\gamma$-softness, we discuss them from the perspective of prolate-oblate shape phase transition. Shape phase transition is an important research topic in nuclear structures \cite{Jolie03t,Casten06,Luo06,Casten07,Wang08,Luo09,Casten09,Casten10,Kotila12,Jolos21,Fortunao21}. More discussions focus on the spherical to deformed-shapes phase transition \cite{Casten10}, and the critical points can be described by exact symmetry, such as the $E(5)$ symmetry for the spherical to $\gamma$-unrelated shape transition \cite{Iachello00,Zhangyu22}, the $X(5)$ symmetry for the spherical to prolate shape transition \cite{Iachello01}, and the $T(5)$, $T(4)$ symmetries for the spherical to $\gamma$-rigid triaxial shape transition \cite{Zhang15,Zhang17}. If spherical vibrational mode is questioned \cite{Heyde16,Garrett18}, such shape phase transition needs further discussions.

Shape phase transitions between different deformations are less discussed because experimental data are sparse. The critical points between different deformations can be described by $Y(5)$ symmetry for the prolate to rigid-triaxiality shape transition \cite{Iachello03}, or the $Z(5)$ symmetry for the prolate-oblate shape transition \cite{Bonatsos04}. In our paper, based on the interesting findings in \cite{Wang22,Fortunato11,Zhang12}, the prolate-oblate shape phase transition is further discussed. This kind of shape phase transition is still not well understood theoretically in IBM.

\begin{figure}[tbh]
\includegraphics[scale=0.27]{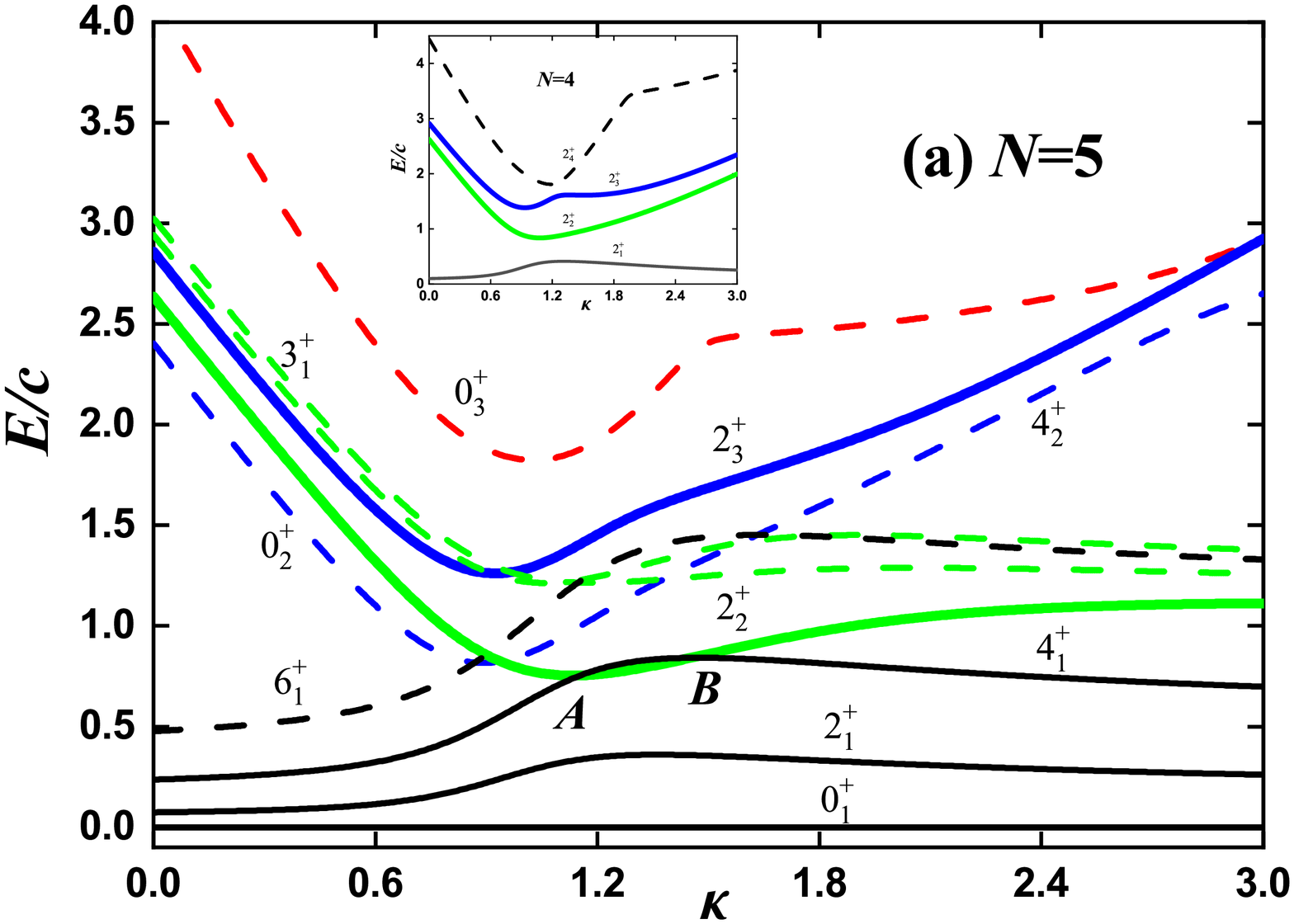}
\includegraphics[scale=0.27]{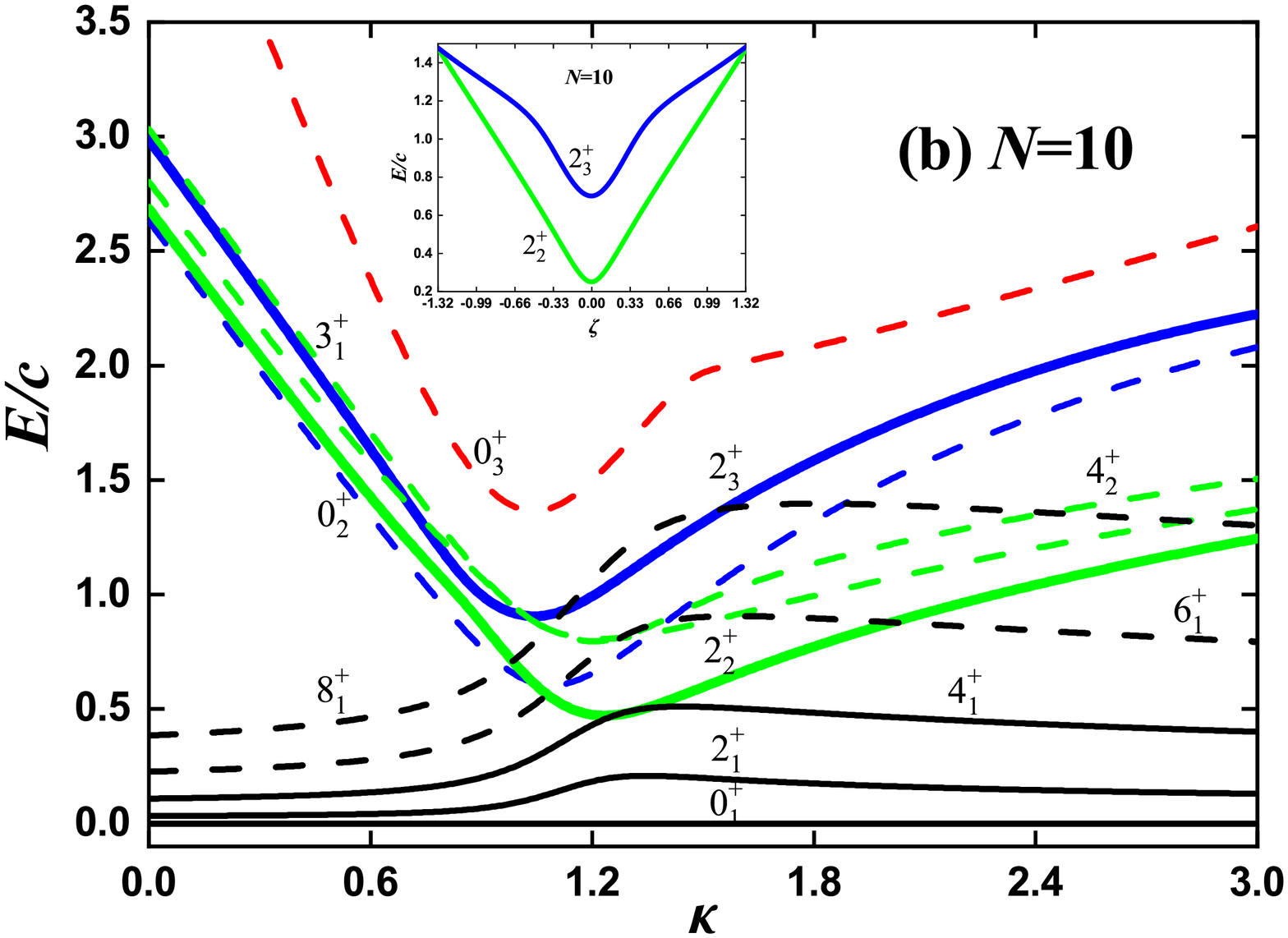}
\includegraphics[scale=0.27]{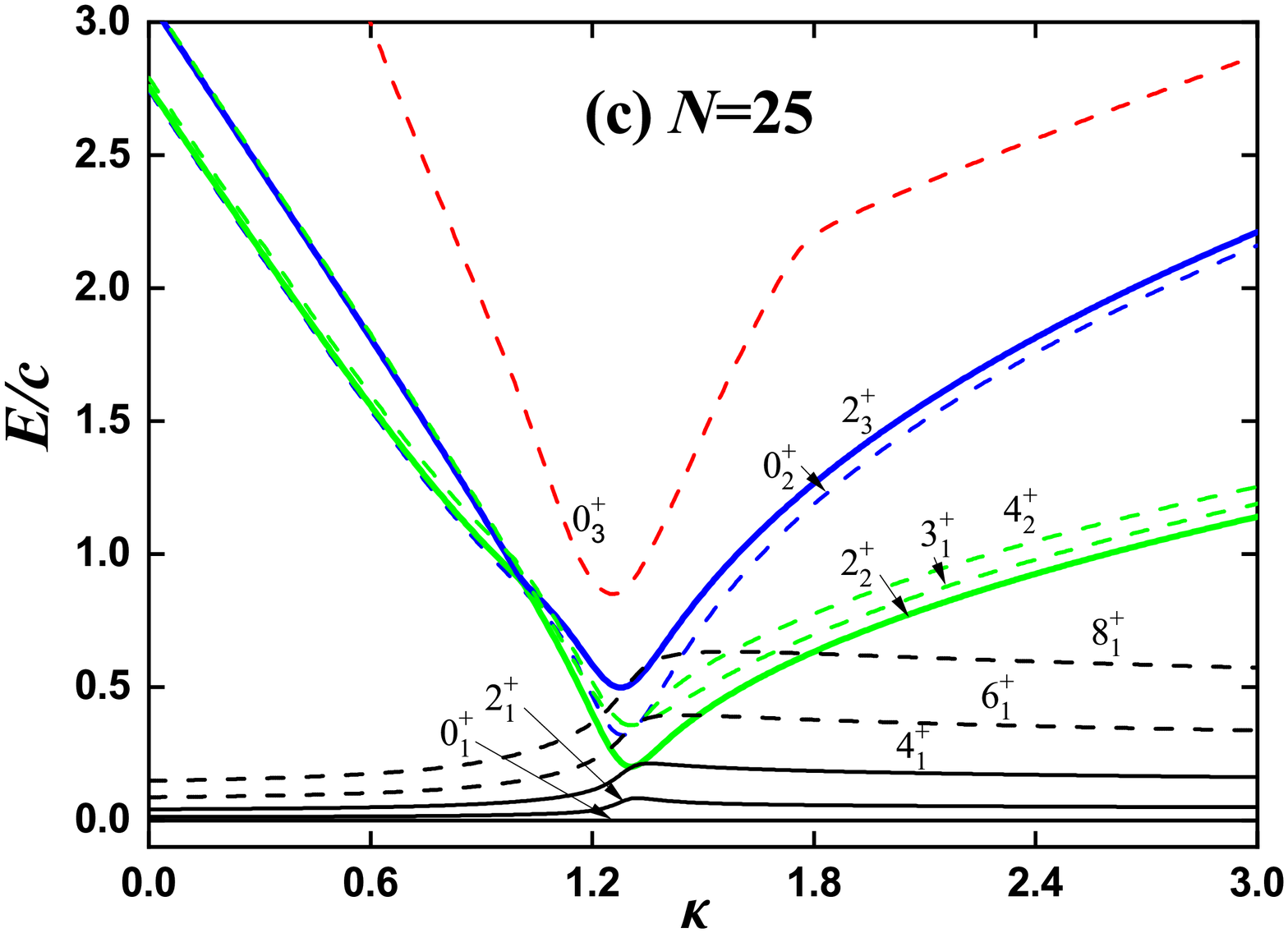}
\caption{The level evolutional behaviors of some low-lying states of the SU3-IBM for (a) $N=5$, (b) $N=10$, and (c) $N=25$. The inset in (a) presents the level evolutional behaviors of some low-lying $2^{+}$ states of the SU3-IBM for $N=4$. The inset in (b) presents the level evolutional behaviors of the $2_{2}^{+}$ and $2_{3}^{+}$ states of the $SU(3)$ limit to the $\overline{SU(3)}$ limit in previous IBM for $N=10$.}
\end{figure}

The conclusion that there exists a phase transition from the prolate shape to the oblate shape in the IBM-1 was first pointed out in the context of catastrophe theory with the coherent state formalism (the large $N$ limit) in Ref. \cite{Castanos96,Castanos98}. Then this prolate-oblate shape transition was numerically studied in detail along the $SU(3)-O(6)-\overline{SU(3)}$ line in the IBM-1, in which the proton-pair and neutron-pair are not distinguished \cite{Jolie01}, see Fig. 2 (a). The experimental data in the Hf-Hg mass region was fitted within the prolate-oblate shape phase transition \cite{jolie03}. The $O(6)$ limit is not only a dynamical symmetry of the $U(6)$ group of the IBM, but also the critical point of the prolate-oblate phase transition \cite{Jolie01}. The prolate shape and the oblate shape are mirror symmetric about the $O(6)$ $\gamma$-soft point from the spectra perspective (detailed discussions on this symmetric feature can be found in \cite{Wang08}).

Starting from the geometric model, a critical point $Z(5)$ symmetry was introduced for the prolate to oblate shape phase transition \cite{Bonatsos04}, which is somewhat different from the $O(6)$ description. In this description, the potential energy is also related to the $\gamma$ geometric variable, and the $O(5)$ symmetry does not hold. $^{194}$Pt is confirmed as the critical nucleus with $Z(5)$ symmetry \cite{Bonatsos04}. Fortunato \emph{et al.} discussed the prolate-oblate shape transition in the large $N$ limit along the blue line in Fig. 1 in the extended cubic-$Q$ IBM, which is also discussed in this paper numerically, see Fig. 2 (c). It was shown that, this prolate-oblate evolution is asymmetric (see the Fig. 12 in \cite{Fortunato11} and the detailed discussions in that paper). The prolate-oblate shape transition was discussed analytically by Zhang \emph{et al.} \cite{Zhang12}, see Fig. 2 (b). These works first show that, shape phase transition from the prolate shape to the oblate shape is not a symmetric evolution, which is in line with the actual evolution of the nuclei in the Hf-Hg region, see Fig. 3. It can be found that, microscopic theories based on energy density functional also support this asymmetric evolution (see Ref. \cite{Nomura11t,Nomura11,Li21} and the detailed discussions in these papers). In addition, the prolte-oblate shape phase transition was also discussed in the proxy-$SU(3)$ model, which highlights the prolate dominance \cite{Bohatsos17,Bohatsos23}. Recently, an important progress was made that microscopic mechanism based on the shell model is revealed for the oblate-prolate shape transition in the Te-Xe-Ba region \cite{Shimizu23}, in which the quadrupole moment is regarded as an important order parameter for the prolate-oblate shape phase transition,   and quasi-$SU(3)$ couplings play a critical role in driving shape evolution and phase transition \cite{Zuker95,Zuker15}. In our paper, the SU3-IBM description along the blue line in Fig. 1 can describe the realistic prolate-oblate shape phase transition in the Hf-Hg region, which further validates the new theory. Fig. 3 presents the experimental quadrupole moments of the first and second $2^{+}$ states, which shows prominent prolate-oblate asymmetry, an abruptness of the shape phase transition and possible crossing phenomenon between the $2_{1}^{+}$ and $2_{2}^{+}$ states. These peculiar features will be explained in the new model. These results show that, oblate shape, $\gamma$-softness and $B(E2)$ anomaly may have a common origin, which is very important for us to understand the evolution of nuclear structure and the emergence of collective behaviors. The significance of the new theory will be discussed in the discussion section.

\section{Hamiltonian}

\begin{figure}[tbh]
\includegraphics[scale=0.27]{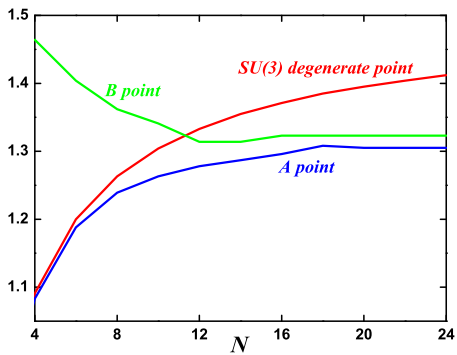}
\caption{The evolutional behaviors of the values of the $A$ point, $B$ point and $SU(3)$ degenerate point when boson number $N$ increases.}
\end{figure}

The Hamiltonian for describing the prolate-oblate shape phase transition has been discussed in \cite{Wang22}, which is
\begin{eqnarray}
\hat{H}=c[(1-\eta) \hat{n}_{d}+\eta (-\frac{\hat{C}_{2}[SU(3)]}{2N}+\kappa \frac{\hat{C}_{3}[SU(3)]}{2N^{2}})],
\end{eqnarray}
where $0\leq \eta \leq 1$, \emph{c} is the total fitting parameter,\emph{ N} is the boson number, $\kappa$ is the coefficient of the cubic interaction, $\hat{n}_{d}$ is the $d$ boson number operator, $\hat{C}_{2}[SU(3)]$ and $\hat{C}_{3}[SU(3)]$ are the second-order and third-order $SU(3)$ Casimir operators separately. The Hamiltonian (1) can be described by the new spherical-prolate-oblate shape triangle in Fig. 1. If $\eta=0$, it describes the spherical shape having harmonic vibration. This term is necessary for it represents the pairing interaction.
If $\eta=1$, the second term presents the $SU(3)$ limit. $-\hat{C}_{2}[SU(3)]$ describes the prolate shape, and $\hat{C}_{3}[SU(3)]$ can give an oblate shape description. For $-\hat{C}_{2}[SU(3)]$, the ground state is the $SU(3)$ irreducible representation $(2N,0)$. For $\hat{C}_{3}[SU(3)]$, the ground state is the $(0,N)$. Different $SU(3)$ irreducible representation corresponds to different quadrupole shapes \cite{Isacker00,zhang14}, which is the reason why the $SU(3)$ limit has various shapes \cite{Zhang12}. Thus the key thing is to make one representation $(\lambda, \mu)$ becomes the ground state \cite{Zhang22}.

The $\overline{SU(3)}$ description of the oblate shape in previous IMB is replaced by the $SU(3)$ third-order interaction is the critical difference in the new SU3-IBM theory. If necessary, other $SU(3)$ higher-order interactions can be introduced, such as $\hat{C}_{2}^{2}[SU(3)]$, $[\hat{L}\times \hat{Q} \times \hat{L}]^{(0)}$ and  $[(\hat{L}\times \hat{Q})^{(1)} \times (\hat{L} \times \hat{Q})^{(1)}]^{(0)}$, where $\hat{Q}$ is the $SU(3)$ quadrupole operator, and $\hat{L}$ is the angular momentum operator. For the description of $B(E2)$ anomaly, these interactions are all important \cite{Zhang22}. Thus the Hamiltonian (1) is the simplest formalism in the SU3-IBM for the study of the new $\gamma$-softness.

\begin{figure}[tbh]
\includegraphics[scale=0.27]{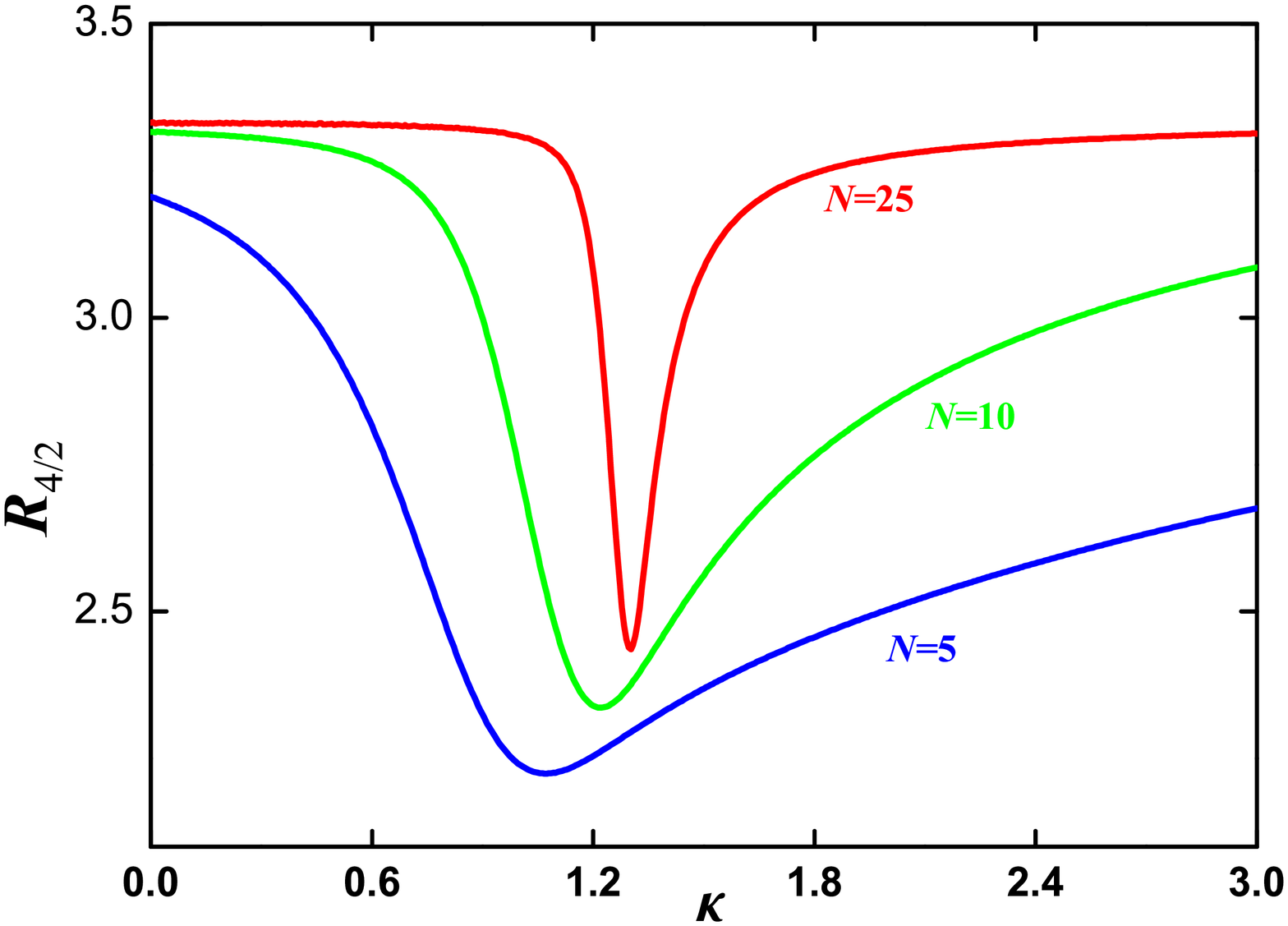}
\includegraphics[scale=0.27]{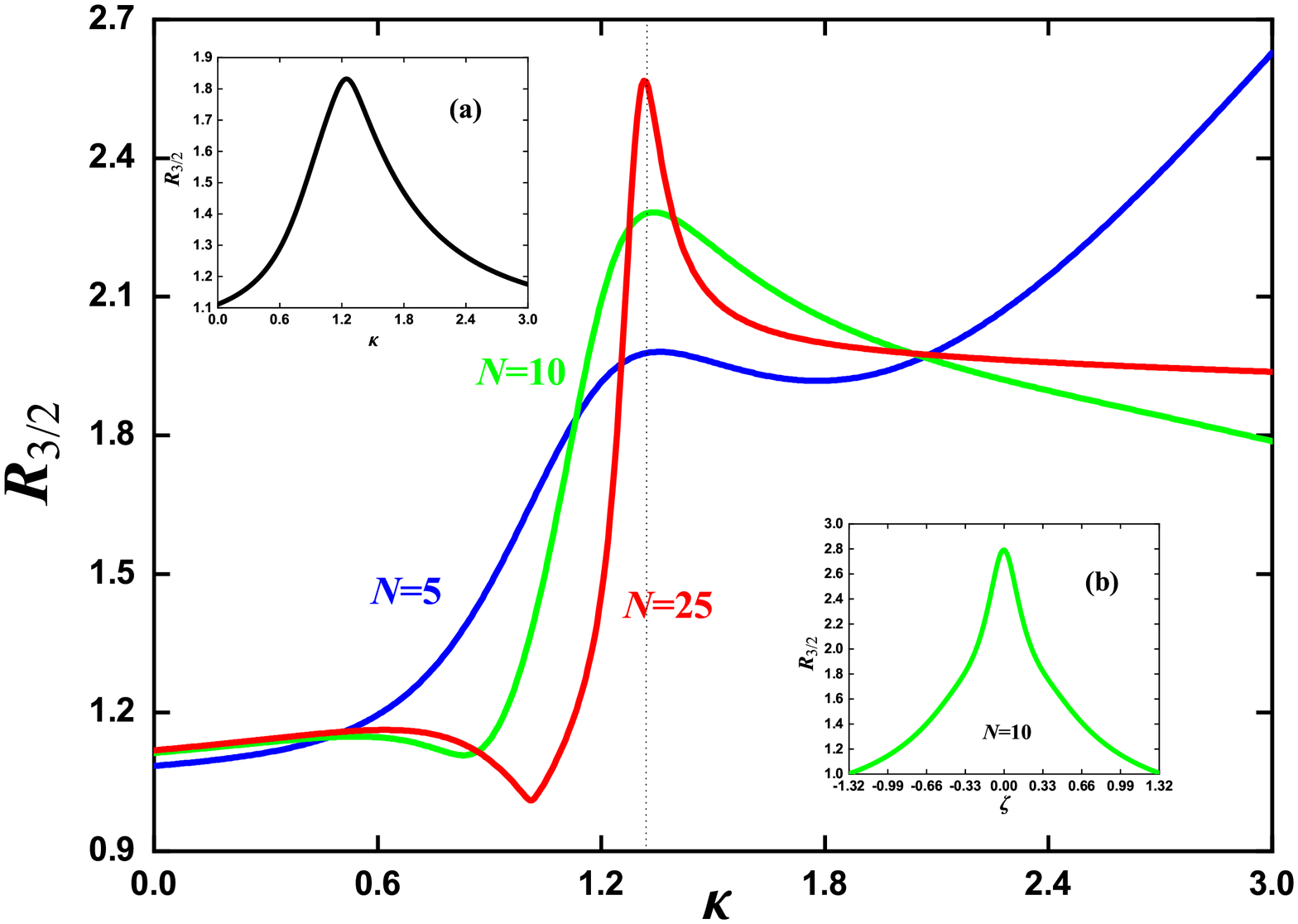}
\caption{The evolutional behaviors of the ratios $R_{4/2}$ and $R_{3/2}$ for $N=5$, $N=10$ and $N=25$. The inset (a) in the below figure presents the evolutional behavior of the $R_{3/2}$ for $N=4$ and the inset (b) presents the evolutional behavior of the $R_{3/2}$ along the $SU(3)-O(6)-\overline{SU(3)}$ line in previous IBM for $N=10$.}
\end{figure}

\begin{figure}[tbh]
\includegraphics[scale=0.25]{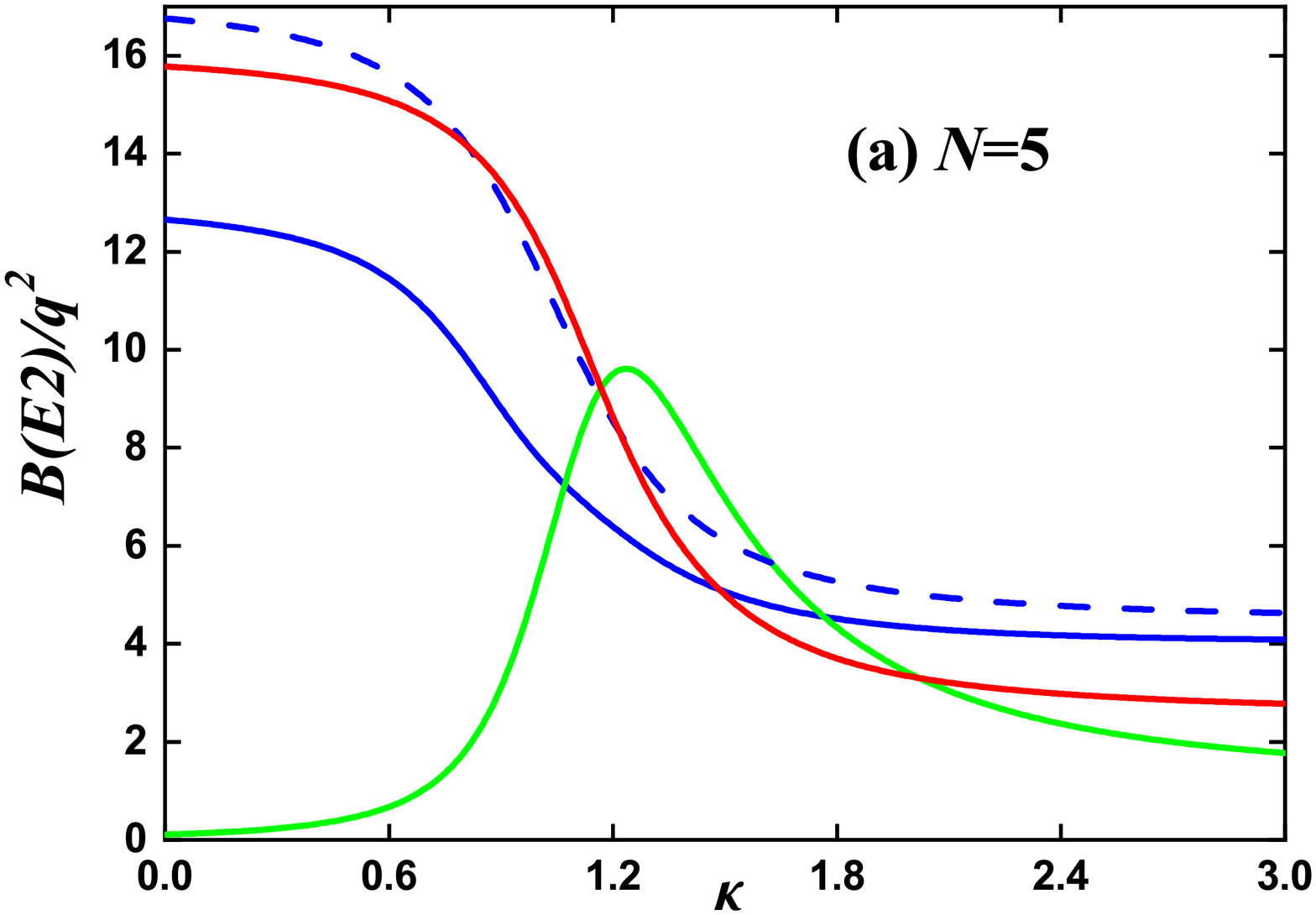}
\includegraphics[scale=0.25]{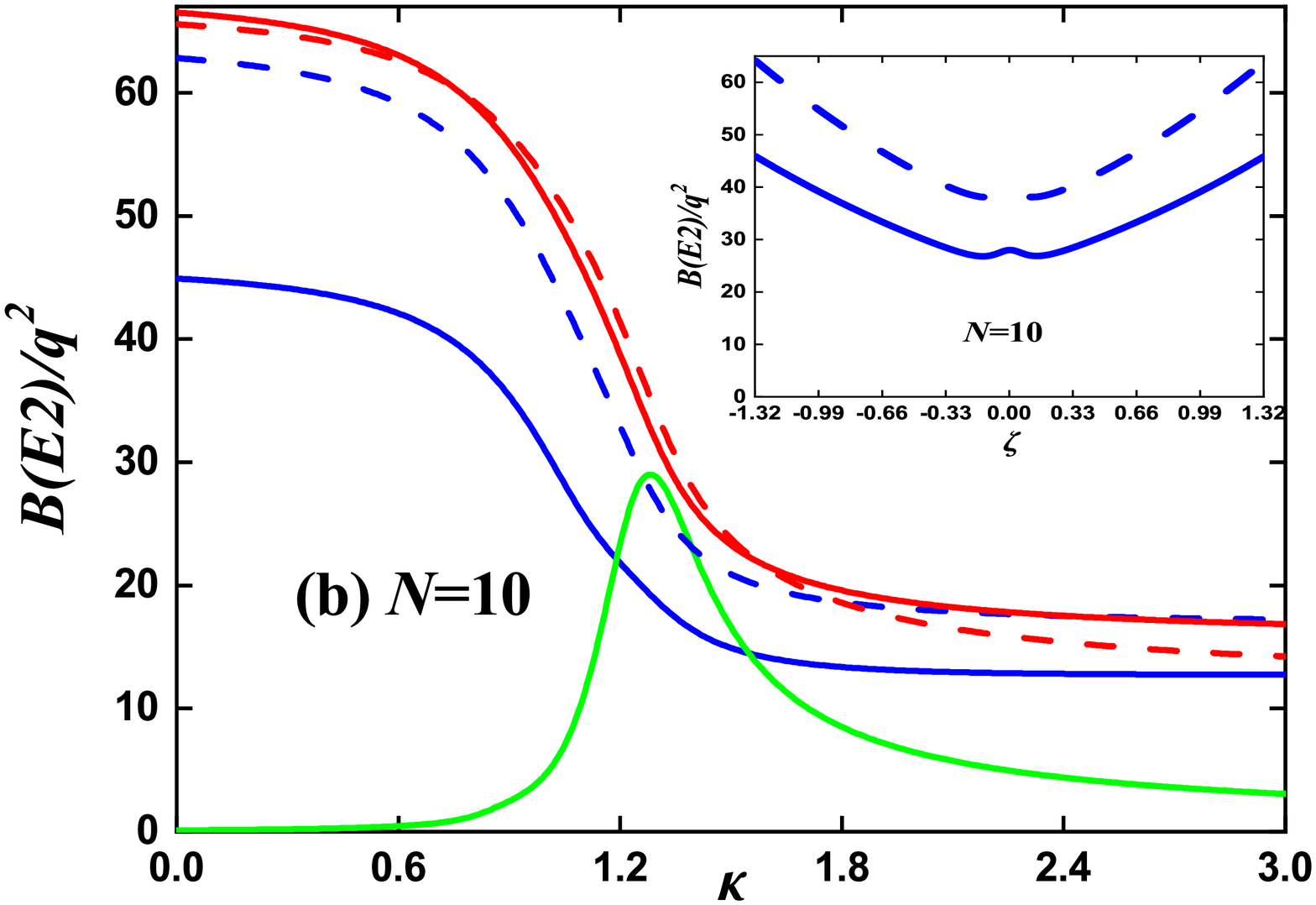}
\includegraphics[scale=0.25]{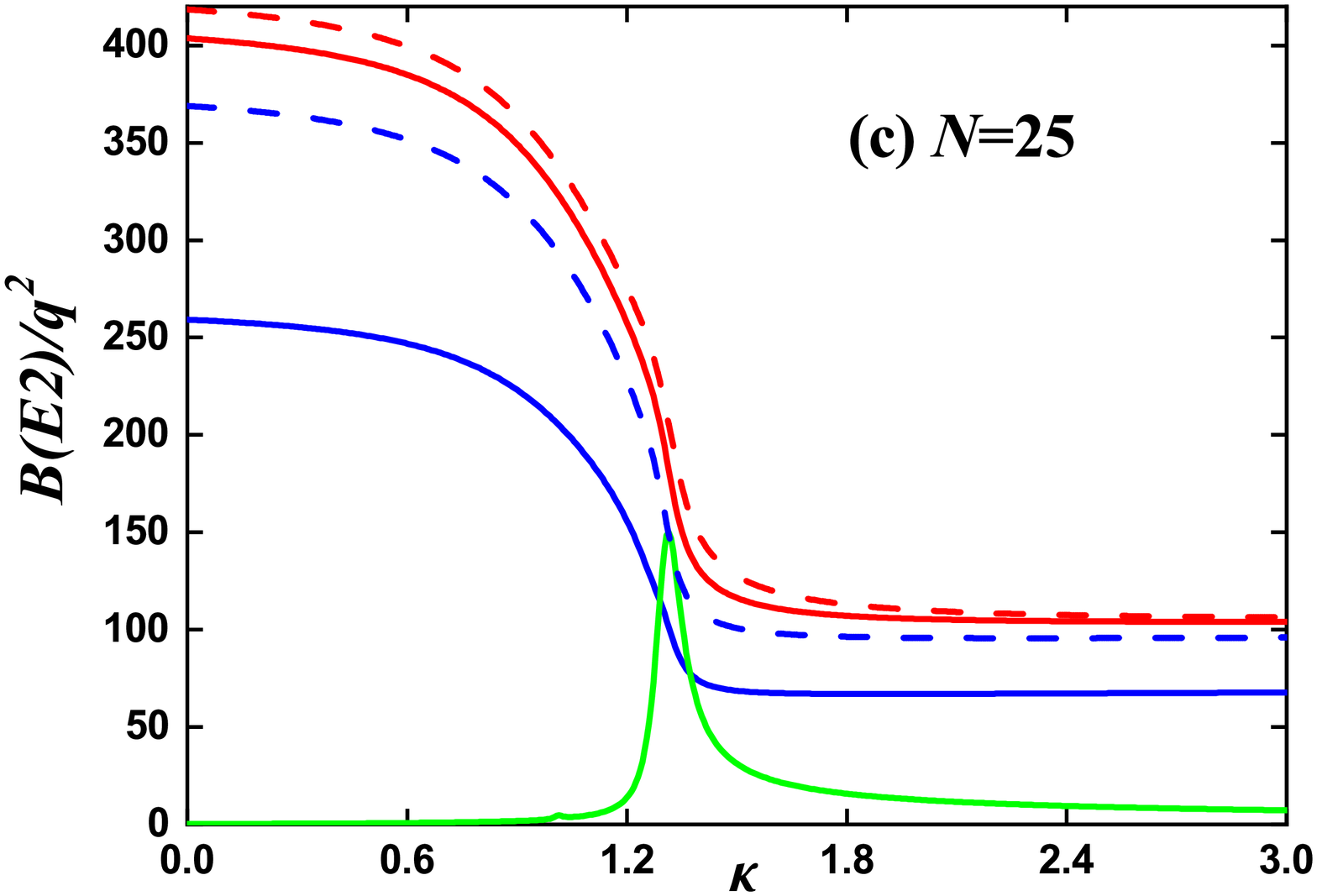}
\caption{The evolutional behaviors of $B(E2; 2_{1}^{+}\rightarrow 0_{1}^{+})$ (blue real line), $B(E2; 4_{1}^{+}\rightarrow 2_{1}^{+})$ (blue dashed line), $B(E2; 6_{1}^{+}\rightarrow 4_{1}^{+})$ (red real line), $B(E2; 8_{1}^{+}\rightarrow 6_{1}^{+})$ (red dashed line), and $B(E2; 2_{2}^{+}\rightarrow 2_{1}^{+})$ (green real line) along the blue line in Fig. 1 for (a) $N=5$, (b) $N=10$ and (c) $N=25$. The inset in (b) presents the evolutional behaviors of $B(E2; 2_{1}^{+}\rightarrow 0_{1}^{+})$, $B(E2; 4_{1}^{+}\rightarrow 2_{1}^{+})$ along the $SU(3)$-$O(6)$-$\overline{SU(3)}$ line in previous IBM for $N=10$.}
\end{figure}

Fortunato \emph{et al.} generalized the simple IBM-1 formalism including a cubic-$Q_{\chi}$ interaction \cite{Fortunato11}, which is
\begin{small}
\begin{equation}
\hat{H}'=c [(1-\eta) \hat{n}_{d}-\frac{\eta}{N} (\hat{Q}_{\chi}\cdot \hat{Q}_{\chi}+ \frac{\kappa_{3}}{N}[\hat{Q}_{\chi}\times \hat{Q}_{\chi}\times \hat{Q}_{\chi}]^{(0)} ) ],
\end{equation}
\end{small}
where $\kappa_{3}$ is the coefficient of the cubic term. $\hat{Q}_{ \chi}=[d^{\dag}\times\tilde{s}+s^{\dag}\times \tilde{d}]^{(2)}+\chi[d^{\dag}\times \tilde{d}]^{(2)} $ is the generalized quadrupole operator, and $-\frac{\sqrt{7}}{2}\leq \chi \leq \frac{\sqrt{7}}{2}$, $\kappa_{3}=\frac{ 2\sqrt{35}\kappa}{9}$.
If $\eta=1$ and $\kappa_{3}=0$, Hamiltonian (2) describes the prolate-oblate shape phase transition from the $SU(3)$ limit ($\chi=-\frac{\sqrt{7}}{2}$) to the $\overline{SU(3)}$ limit ($\chi=\frac{\sqrt{7}}{2}$) via the $O(6)$ critical point ($\chi=0$), see Fig. 2 (a).

For $\chi=-\frac{\sqrt{7}}{2}$, the quadrupole second or third-order interactions can be related with the $SU(3)$ two Casimir operators as following
\begin{equation}
\hat{C}_{2}[SU(3)]=2\hat{Q}\cdot \hat{Q}+\frac{3}{4} \hat{L}\cdot \hat{L},
\end{equation}
\begin{small}
\begin{equation}
\hat{C}_{3}[SU(3)]=-\frac{4}{9}\sqrt{35}[\hat{Q}\times \hat{Q} \times \hat{Q}]^{(0)}-\frac{\sqrt{15}}{2}[\hat{L}\times \hat{Q} \times \hat{L}]^{(0)}.
\end{equation}
\end{small}
For a certain $SU(3)$ irrep $(\lambda,\mu)$, the eigenvalues of the two Casimir operators under the group chain $U(6)\supset SU(3) \supset O(3)$ can be expressed as
\begin{equation}
\langle \hat{C}_{2}[SU(3)]\rangle=\lambda^{2}+\mu^{2}+\lambda \mu+3\lambda+3\mu,
\end{equation}
\begin{equation}
\langle \hat{C}_{3}[SU(3)]\rangle=\frac{1}{9}(\lambda-\mu)(2\lambda+\mu+3 ) (\lambda+2\mu+3 ).
\end{equation}
If $\kappa_{c}=\frac{3N}{2N+3}$, the second term in Hamiltonian (1) describes the $SU(3)$ degenerate point. It should be noticed that the location of the $SU(3)$ degenerate point $\kappa_{c}$ is relevant to the boson number $N$. At this degenerate point, the $SU(3)$ irreducible representations satisfying the condition $\lambda+2\mu=2N$ are all degenerate. Thus for $\chi=-\frac{\sqrt{7}}{2}$, the Hamiltonian (2) can have the same $0^{+}$ states as the Hamiltonian (1), but has different energies for states with angular momentum $L> 0$. The Hamiltonian (1) can show more regular patterns, such as $O(5)$ partial dynamic symmetry, which offers a new $\gamma$-soft rotational mode \cite{Wang22}. Moreover, the three-body interaction $[\hat{L}\times \hat{Q} \times \hat{L}]^{(0)}$ in Equ. (4)  can also be naturally introduced into the Hamiltonian (1), which is vital for the explanation of $B(E2)$ anomaly \cite{Wang20}.

When $\eta=1$, and $\kappa$ increases, the Hamiltonian (1) can describe a sudden change of the shapes for the ground state \cite{Zhang12}, see Fig. 2 (b). The phase transition point is
$\kappa_{c}=\frac{3N}{2N+3}$. However, this asymmetric evolution path (the green line in Fig. 1) is a degraded description for the prolate-oblate shape phase transition, because the $\gamma$-softness vanishes. For a realistic description, the $\hat{n}_{d}$ interaction must be included. In Ref. \cite{Fortunato11}, Fortunato \emph{et al.} studied the evolution case from the prolate shape to the oblate shape in the large $N$ limit based on the coherent state formalism (see the Fig. 12 in that paper), which corresponds to the blue line ($\eta=0.5$) in Fig. 1, see also Fig. 2 (c). It is clearly shown that, this prolate-oblate shape transition is not symmetric, and experiences a $\gamma$-soft region with a critical point from the prolate-biased shape to the oblate-biased shape. Our paper will provide a detailed numerical investigation along this evolution path with parameter $\kappa$.

\begin{figure}[tbh]
\includegraphics[scale=0.27]{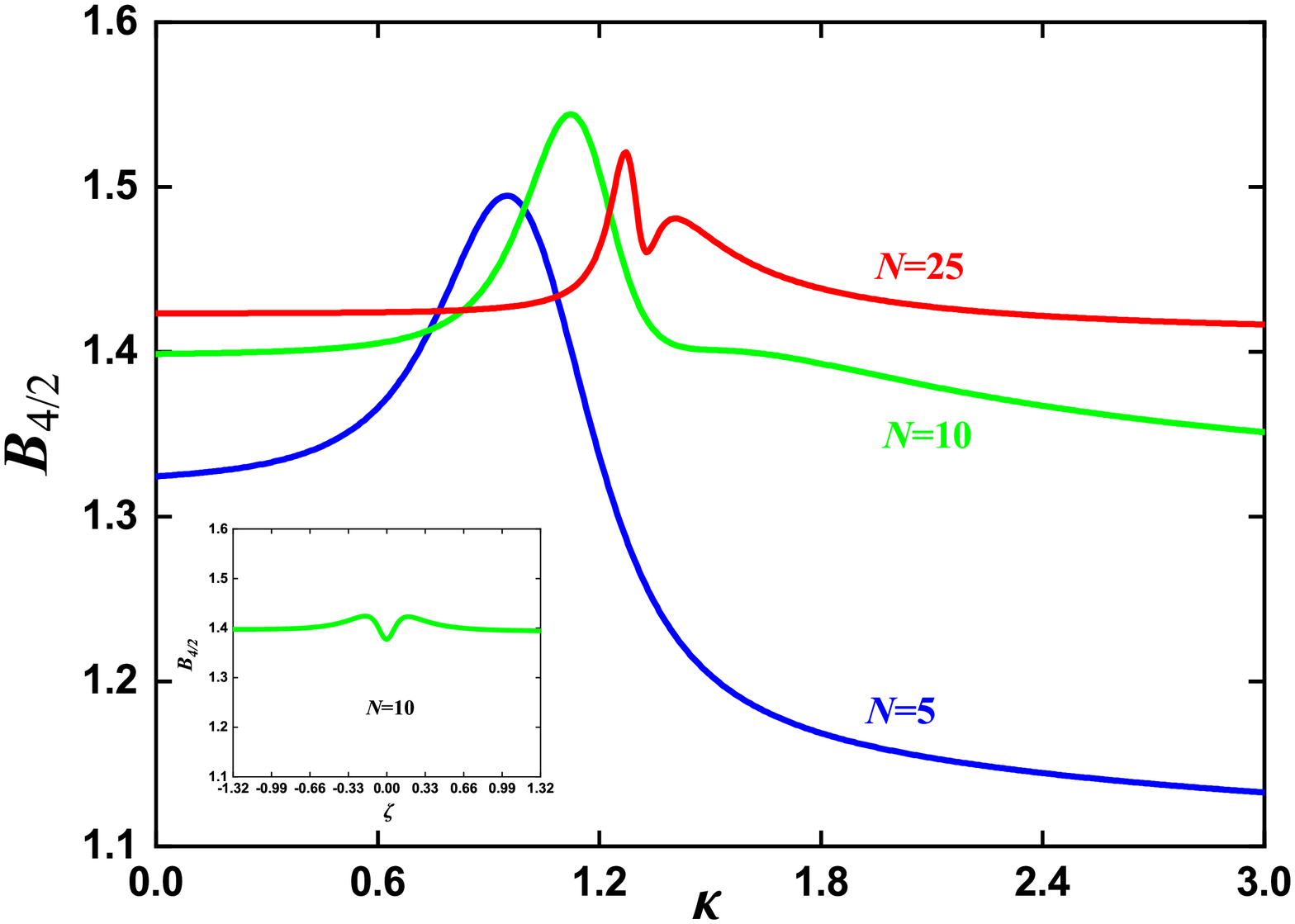}
\caption{The evolutional behaviors of the ratio $B_{4/2}$ for $N=5$, $N=10$ and $N=25$. The inset presents the evolutional behaviors of the ratio $B_{4/2}$  along the $SU(3)$-$O(6)$-$\overline{SU(3)}$ line in previous IBM for $N=10$.}
\end{figure}

\section{Prolate-oblate shape phase transition}

Prolate-oblate shape phase transition is more conducive to understand the interactions between nucleons and quadrupole deformations in nuclear structure \cite{Shimizu23}. Ref. \cite{Fortunato11} showed that, there is no a phase transition point from the prolate shape to the oblate shape along the blue line in Fig. 1. There is actually a narrow region $1.333 \leq \kappa \leq 1.362$ with rigid triaxiality having shallow $\gamma$ potential. In the large $N$ limit, if $\kappa < 1.333$, it corresponds to the prolate shape, while if $\kappa > 1.362$, it corresponds to the oblate shape. If $1.333 \leq \kappa \leq 1.362$, it corresponds to a triaxial shape. This evolution case is somewhat different from the one in previous IBM with the $O(6)$ limit as the critical point \cite{Wang08}. In the new theory, the belief that triaxiality comes from the competition between the prolate shape and the oblate shape is better illustrated.

Level evolutions of some low-lying states for (a) $N=5$, (b) $N=10$, and (c) $N=25$ are shown in Fig. 4 when $\kappa$ increases from 0 to 3.  For quadrupole deformation, $0_{1}^{+}$, $2_{1}^{+}$, $4_{1}^{+}$, $6_{1}^{+}$, $8_{1}^{+}$ states in the ground band, $2_{2}^{+}$, $3_{1}^{+}$, $4_{2}^{+}$ states in the $\gamma$ band, $0_{2}^{+}$, $2_{3}^{+}$ states in the $\beta$ band, and the $0_{3}^{+}$ state are presented. The behavior of the prolate-oblate shape phase transition is quite obvious even for $N=10$. The left side is the rotational spectra of a prolate shape, for which the $\gamma$ band and the $\beta$ band are in close proximity. The right side is the oblate rotational spectra, for which the bandhead $2_{2}^{+}$ state of the $\gamma$ band is lower than the $0_{2}^{+}$ state of the $\beta$ band. The oblate spectra is obviously different from the prolate one. The transitional part presents the new $\gamma$-soft spectra \cite{Wang22,Wang23}. For $N=25$, the $\gamma$-soft region have been reduced to a very short length. Thus the emergent $\gamma$-softness is also a finite-$N$ effect.

To distinguish between the prolate shape and the oblate shape, it should be noted that, the position relationship between the $2_{2}^{+}$ and $2_{3}^{+}$ states is a critical indicator. For small $N$, the value of the position of the minimum energy value of the $2_{3}^{+}$ state is smaller than the one of the $2_{2}^{+}$ state. For $N=10$, the two positions are $\kappa=1.044$ and $\kappa=1.23$. For the $SU(3)-O(6)-\overline{SU(3)}$ line in previous IBM, the two positions are the same at the $O(6)$ critical point (see the inset in Fig. 4 (b)). At the prolate side, the $2_{2}^{+}$ and $2_{3}^{+}$ states are very close together ($\kappa<1.044$ for $N=10$). At the oblate side, the two states are far apart for $N\geq 5$, while not so for $N=4$ (see the inset in Fig. 4 (a)). For $N=25$, it is clear that, the $2_{2}^{+}$ and $2_{3}^{+}$ states crossover with each other approximately at the prolate side. The feature (energy repulsion) can emerge even for $N=10$. This class of signature is particularly useful for identifying specific shape quantum phase transitions, for example the crossover of the $0_{2}^{+}$ and $0_{3}^{+}$ states in previous IBM \cite{Wang08} and in the SU3-IBM newly found \cite{Zhou23}. In this paper, the overall evolutionary behaviors of the two $2_{2}^{+}$ and $2_{3}^{+}$ states are very critical for confirming the rationality of the SU3-IBM.

Two locations of the phase transition region has been discussed \cite{Wang22,Wang23}. The $2_{2}^{+}$ and $4_{1}^{+}$ states have two crossing points. The left point is called $A$ point, while the right one is called $B$ point, see Fig. 4 (a).  The spectra of $A$ point have $O(5)$ partial dynamical symmetry, which seems to be able to fit the normal states of the Cd isotopes \cite{Wang22}. This partial dynamical symmetry is different from the one discussed in \cite{Leviatan11}, and the origin of the new symmetry is still unknown. The spectra of the $B$ point can be used to explain the excitations in $^{196}$Pt ($\hat{C}_{2}^{2}[SU(3)]$ is necessary for the realistic description) \cite{Wang23}, which is a typical $\gamma$-soft nucleus. It should be noticed that, for small $N$, the spectra of the oblate shape are somewhat similar to the $\gamma$-soft ones. This is the possible reason why the oblate nucleus $^{198}$Hg is usually regarded as soft triaxial \cite{Fortunato19}.

\begin{figure}[tbh]
\includegraphics[scale=0.27]{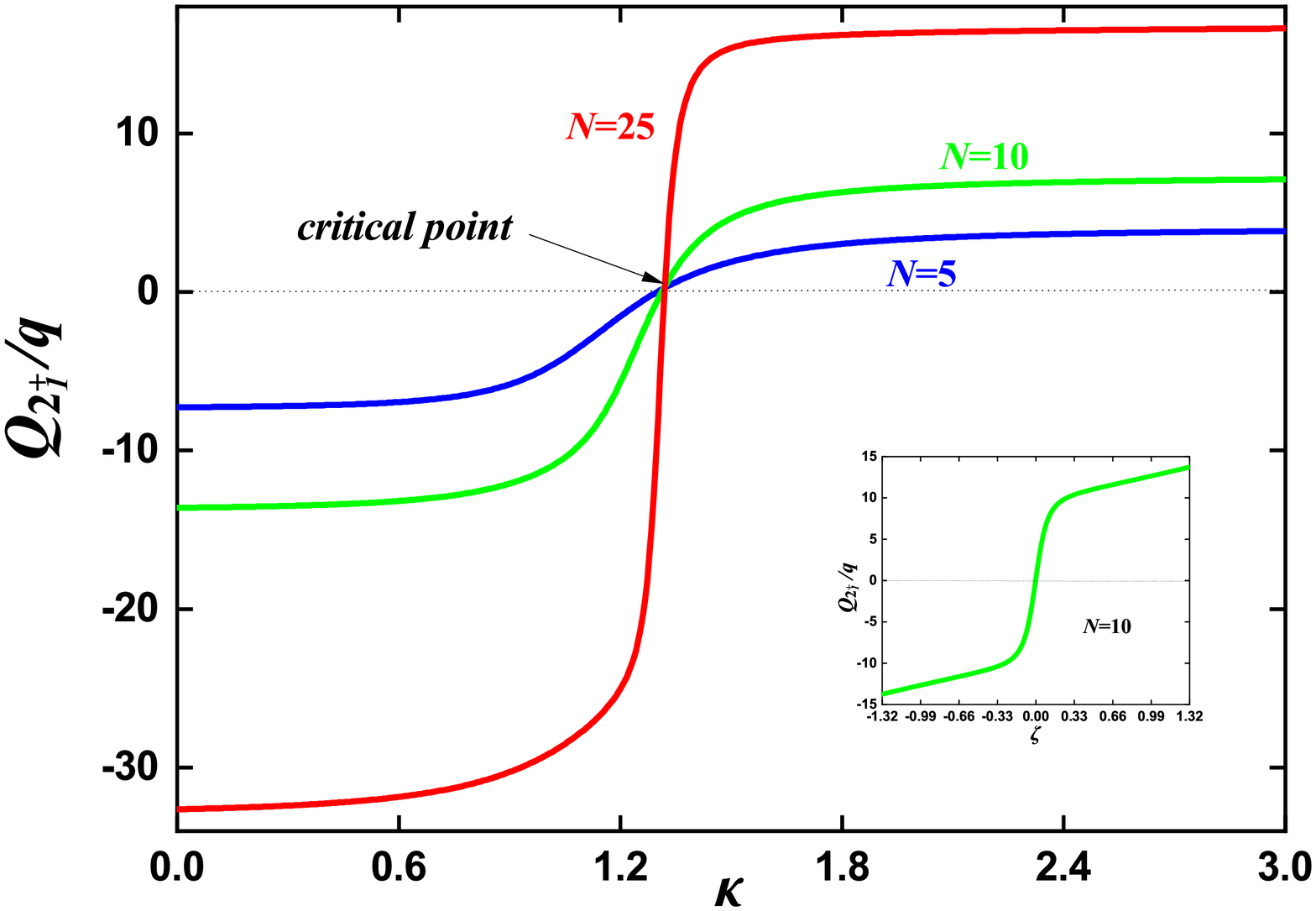}
\includegraphics[scale=0.27]{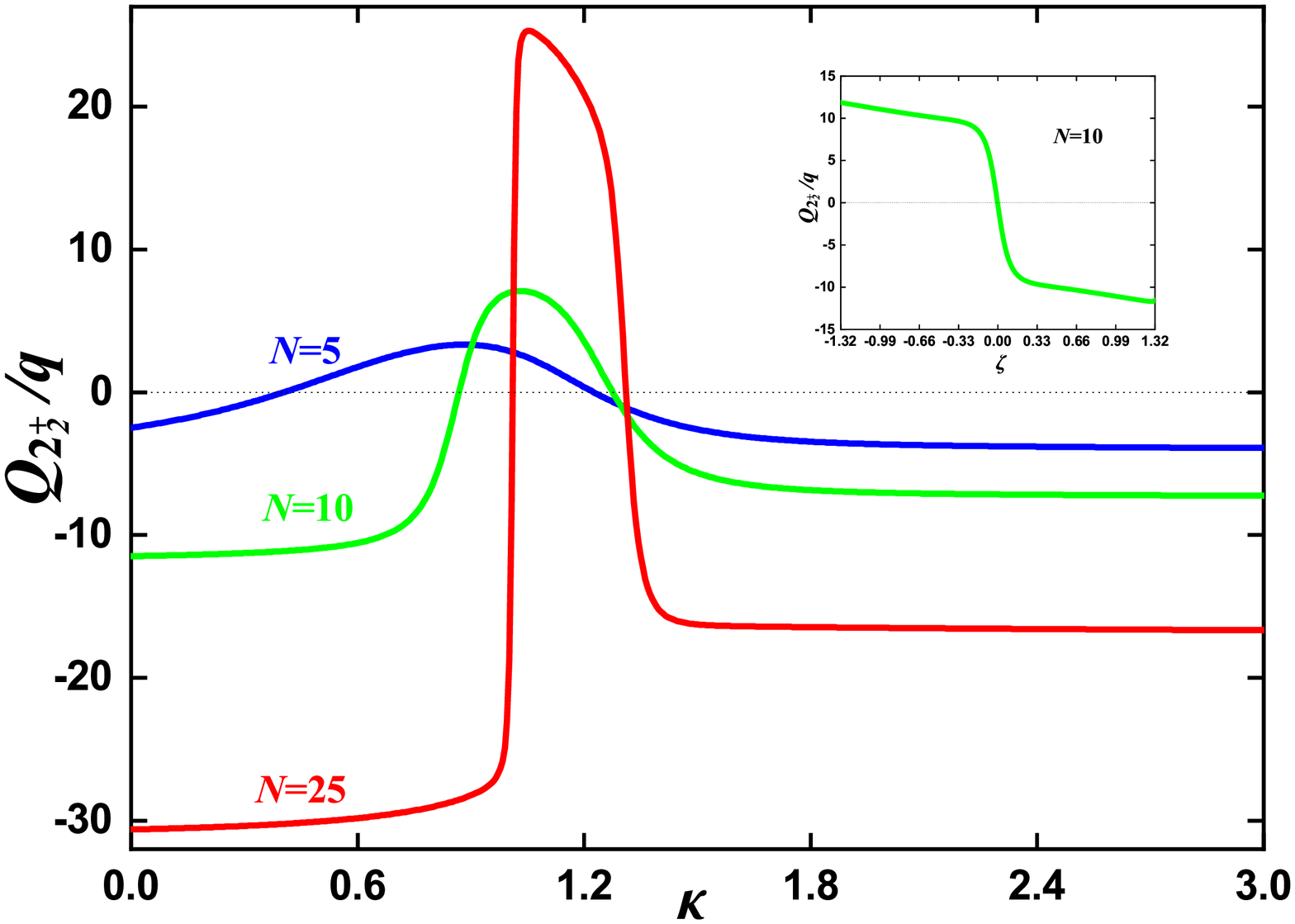}
\includegraphics[scale=0.27]{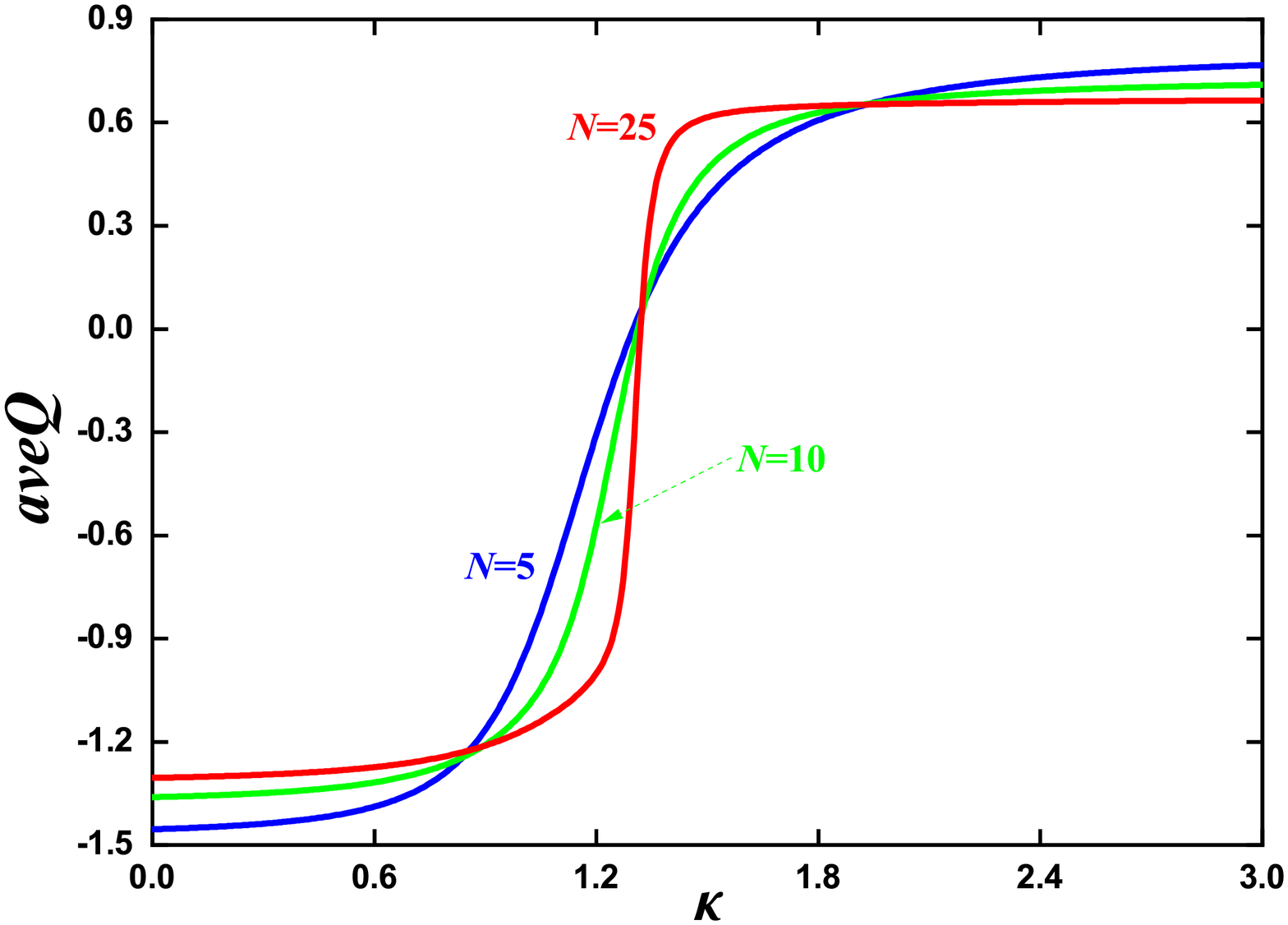}
\caption{The evolutional behaviors of (a) the quadrupole moment $Q_{2_{1}^{+}}$ of the $2^{+}_{1}$ state, (b) the quadrupole moment $Q_{2_{2}^{+}}$ of the $2^{+}_{2}$ state, and (c) the order parameter $ave$Q along the blue line in Fig. 1 for $N=5$, $N=10$ and $N=25$. The insets in (a) and (b) presents the evolutional behaviors of $Q_{2_{1}^{+}}$  and $Q_{2_{2}^{+}}$ along the $SU(3)$-$O(6)$-$\overline{SU(3)}$ line in previous IBM for $N=10$.}
 \end{figure}

The $SU(3)$ degenerate point is a discontinuity point from the prolate shape to the oblate shape. For the green line in Fig. 1 or in Fig. 2 (b), the prolate-oblate shape phase transition is abrupt at the $SU(3)$ degenerate point even for finite $N$ (no $\gamma$-softness). In the large $N$ limit, the $A$ point is the critical point from the prolate shape to the rigid triaxial shape, and the $B$ point is the critical point from the rigid triaxial shape to the oblate shape. Thus the three points are located differently from each other. Fig. 3 presents evolutional behaviors of the values of the $A$ point, $B$ point and $SU(3)$ degenerate point when $N$ increases from 4 to 24. For small $N$, the location of the $A$ point is near the value of the $SU(3)$ degenerate point. However, when $N$ becomes larger, the distance between the two values becomes larger too. Thus the green line in Fig. 1 between the $U(5)$ limit and the $SU(3)$ degenerate point is actually a curve \cite{Fortunato11}, but for small $N$, it is approximately a straight line. A possible relationship between the green curve and the variables $N$ and $\kappa$ in the large $N$ case will be discussed for further understanding the $O(5)$ partial dynamical symmetry. For large $N$, the deviation of the values 1.305 and 1.323 of the $A$, $B$ points from the critical values 1.333 and 1.362 in the large $N$ limit  may come from numerical errors.

For studying shape phase transition, various order parameters are often explored. Especially in the spherical-deformed shape transition, symmetry breaking can occur, and some order parameters can present the deformation. For the prolate-oblate shape phase transition, these order parameters reflect the different degree of deformation. A basic order parameter in nuclear structure is the energy ratio $R_{4/2}=E_{4_{1}^{+}}/E_{2_{1}^{+}}$ of the first $4^{+}$ state and the first $2^{+}$ state. If $R_{4/2}\approx 2$, it is usually regarded as a marker of the spherical shape. If $R_{4/2}\approx 10/3$, this is an indicator for rotational mode of the ellipsoidal shape (prolate or oblate). If $R_{4/2}\approx 2.5$, it could mean a $\gamma$-soft nucleus. Fig. 4 (a) presents the evolutional behaviors of the order parameter $R_{4/2}$ for $N=5$, $N=10$ and $N=25$. For large $N$, previous conclusions still holds. For previous $SU(3)$-$O(6)$-$\overline{SU(3)}$ description, these conclusions are true for any $N$ \cite{Wang08}. For small $N$ in the new model, it is shown that, the evolutional behavior is not symmetric. For the new $\gamma$-softness, it may decrease to 2.3. For the oblate side, it is near 2.7 (not 10/3).

 Because understanding the $2_{2}^{+}$ and $2_{3}^{+}$ states are vital for the prolate-oblate shape phase transition, the energy ratio $R_{3/2}=E_{2_{3}^{+}}/E_{2_{2}^{+}}$ of the two state is also studied in Fig. 4 (b) for $N=5$, $N=10$, $N=5$. Compared with the inset (b) presenting the evolutional behavior of the $R_{3/2}$ along the $SU(3)-O(6)-\overline{SU(3)}$ line in previous IBM for $N=10$, the asymmetry between the prolate side and the oblate side is obvious.  The inset (a) presents the evolutional behavior of the $R_{3/2}$ for $N=4$. There are two key observations. For $N=10$, this value has a rapid increase at the prolate side of the critical point (dotted line) from 1.1 to 2.3. For $N=25$, this sudden change can be very obvious. At the oblate side, for $N\geq 5$, this value is around 2.0, but for $N=4$, it is nearly 1.2.

\section{$B(E2)$ values and quadrupole moment}

Recently, it has been realized that energy spectra alone does not give an accurate judgment of the shape of a nucleus, despite it is a useful starting point for the question. Reduced transition probabilities $B(E2)$ values and other empirical observable quantities are requisite, which relies on the question discussed. In the $B(E2)$ anomaly \cite{168Os,166W,172Pt,170Os,114Xe,114Te,74Zn,50Cr}, the energy spectra can have the same features as the normal experiences. However, the ratio of reduced transition probabilities $B_{4/2}$ within the yrast band can suddenly become much smaller than 1, while nearby nuclei may be perfectly normal with ratio larger than 1. In these nuclei, the energy ratio $E_{4/2}$ is not smaller than 2, thus they show a collective excitation mode. Usually, these nuclei have $\gamma$-soft properties.

\begin{figure}[tbh]
\includegraphics[scale=0.27]{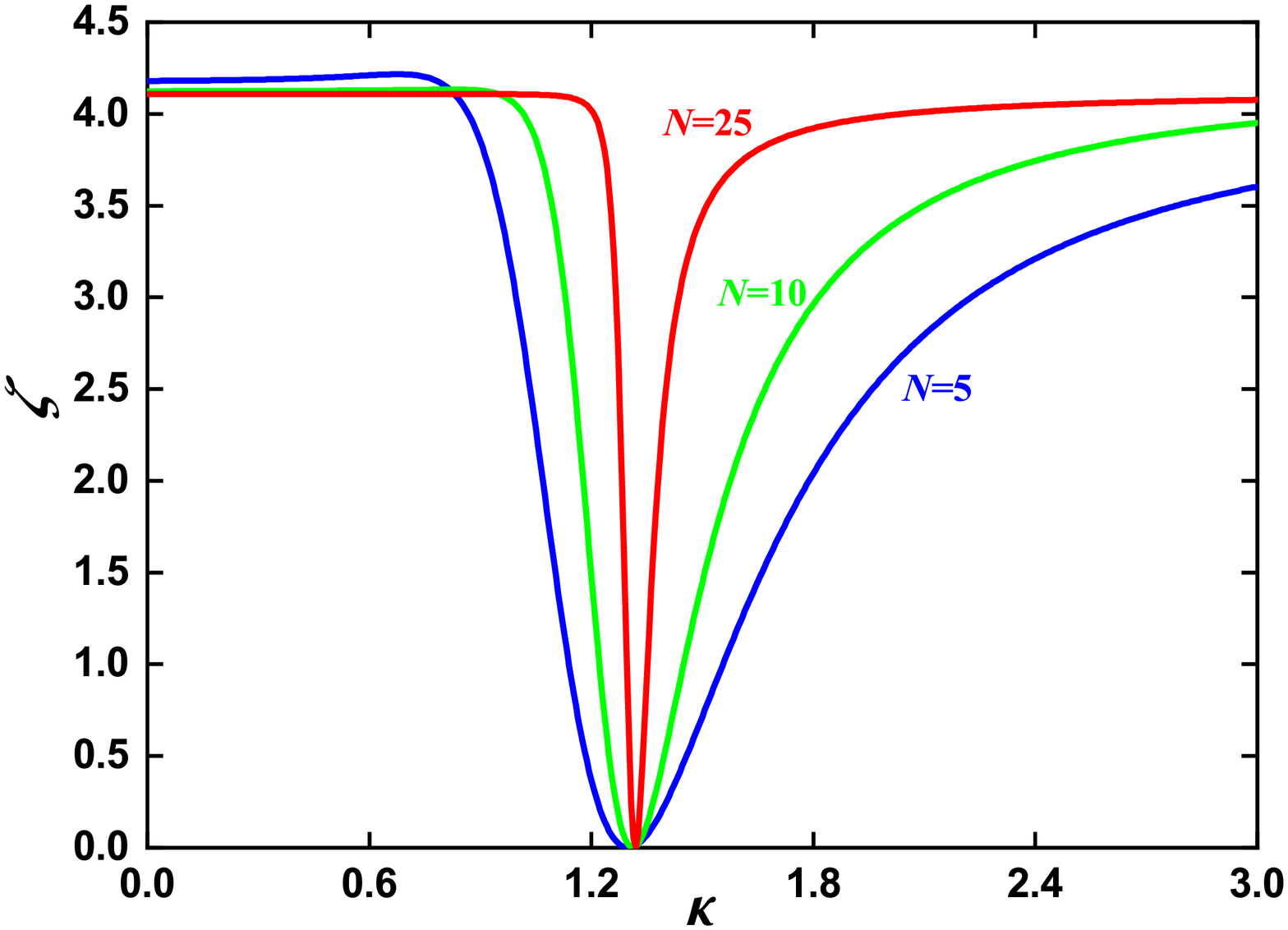}
\caption{The evolutional behaviors of the order parameter $\zeta$ along the blue line in Fig. 1 for $N=5$, $N=10$ and $N=25$.}
\end{figure}

In the spherical nucleus puzzle \cite{Garrett08,Garrett10,Heyde11,Garrett12,Batchelder12,Heyde16,Garrett18,Garrett19,Garrett20}, the spectra seems the spherical vibrational mode \cite{Wang22,Wang23}, but the $B(E2)$ values do not support this conclusion. Ref. \cite{Batchelder12} found that, the $0^{+}$ state near the three-phonon level may not exist. Thus the spectra may be a $\gamma$-soft one \cite{Garrett12}, which is not found in the common studies of modern nuclear structure. This experimental finding is very profound and could even change our understanding about the evolution of nuclear structure \cite{Garrett10,Heyde16,Garrett18,Garrett19}. Understanding the origin of $\gamma$-softness in realistic nuclei may be even more complicated.

\begin{figure}[tbh]
\includegraphics[scale=0.27]{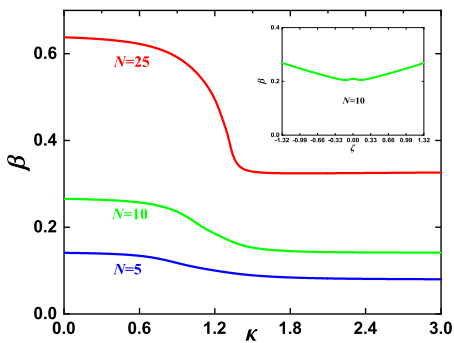}
\caption{The evolutional behaviors of the order parameter $\beta$ along the blue line in Fig. 1 for $N=5$, $N=10$ and $N=25$. The inset presents the evolutional behaviors of $\beta$ along the $SU(3)$-$O(6)$-$\overline{SU(3)}$ line in previous IBM for $N=10$.}
\end{figure}

For a better understanding of the prolate-oblate shape transition in the SU3-model, the evolutional behaviors of $B(E2)$ values of some low-lying states should be studied. The operator is defined as
\begin{equation}
\hat{T}(E2)=q\hat{Q},
\end{equation}
where $q$ is the boson effective charge. The evolutional behaviors of the values $B(E2; 2_{1}^{+}\rightarrow 0_{1}^{+})$ , $B(E2; 4_{1}^{+}\rightarrow 2_{1}^{+})$, $B(E2; 6_{1}^{+}\rightarrow 4_{1}^{+})$, $B(E2; 8_{1}^{+}\rightarrow 6_{1}^{+})$, and $B(E2; 2_{2}^{+}\rightarrow 2_{1}^{+})$ along the blue line in Fig. 1 for (a) $N=5$, (b) $N=10$ and (c) $N=25$ are shown in Fig. 7. The phase transitional behaviors are obvious, and not symmetric like the inset in Fig. 7 (b) along the $SU(3)-O(6)-\overline{SU(3)}$ line in previous IBM for $N=10$. The $B(E2)$ values within the yrast band of the oblate side are smaller than the ones of the prolate side. For $B(E2; 2_{2}^{+}\rightarrow 2_{1}^{+})$, it is small for both the prolate shape and the oblate shape for the $2_{1}^{+}$ state and the $2_{2}^{+}$ state locate at the ground band and the $\gamma$ band individually, and can be comparable to the value of $B(E2; 4_{1}^{+}\rightarrow 2_{1}^{+})$ in the $\gamma$-soft region, which results from the $O(5)$ partial dynamical symmetry.

Order parameter $B_{4/2}$ is also important in shape phase transition. In $B(E2)$ anomaly, it is the key quantity. Fig. 8 presents the evolutional behaviors of the ratio $B_{4/2}$ for $N=5$, $N=10$ and $N=25$. For the prolate-oblate shape transition, the changes of the value are not so prominent for large $N$. For small $N$, the asymmetric shape is still clear. At the prolate side, it is around 1.4, while at the oblate side, it is nearly 1.1. For the $SU(3)$-$O(6)$-$\overline{SU(3)}$ transitions in previous IBM in the inset, it is almost 1.4.

To distinguish between the prolate shape and the oblate shape, the quadrupole moment $Q_{2_{1}^{+}}$ of the $2^{+}_{1}$ state is a key physical quantity. If the deformation is prolate or prolate-biased $\gamma$-soft, the value of this quantity is negative while it is positive for the oblate shape or the oblate-biased $\gamma$-soft. It should be noticed that, prolate (or oblate)-biased $\gamma$-softness does not exist in previous IBM. In the SU3-IBM, the regions from $A$ point to the $B$-point are all $\gamma$-soft, while for previous IBM, only the $O(6)$ limit is so. If the fourth-order interaction $\hat{C}_{2}^{2}[SU(3)]$ is added, this $\gamma$-soft region may be further enlarged (this will be discussed in following papers). Quadrupole moments of the low-lying states are vital for understanding the nuclear structure, but they are difficult to obtain experimentally. Fig. 9 (a) presents the evolutional behaviors of the quadrupole moment of the $2^{+}_{1}$ state along the blue line in Fig. 1 for $N=5$, $N=10$ and $N=25$. It is clearly shown that, the critical point from the prolate-biased to the oblate-biased in the $\gamma$-soft region is near 1.314, which is between the values 1.305 and 1.323 of the $A$, $B$ points. The abrupt change of the quadrupole moment value for the prolate-oblate shape transition is obvious near the critical point. The insets in (a) presents the evolutional behaviors of $Q_{2_{1}^{+}}$  along the $SU(3)$-$O(6)$-$\overline{SU(3)}$ line in previous IBM for $N=10$. It is obvious that, in the new model, the absolute values of the prolate side and the oblate side are not the same, and the value at the prolate side is near twice the one at the oblate side for given $N$. In previous IBM, they are the same.

Fig. 9 (b) presents the evolutional behaviors of the quadrupole moment $Q_{2_{2}^{+}}$ of the $2^{+}_{2}$ state along the blue line in Fig. 1 for $N=5$, $N=10$ and $N=25$. Its performance is very different from the one in previous IBM in the inset. This behavior results from the level repulsion between the $2_{2}^{+}$ and $2_{3}^{+}$ states.

In order to better reflect the asymmetry of the evolutional behavior, the order parameter \emph{aveQ} is introduced
\begin{equation}
aveQ=Q_{2_{1}^{+}}/(qN).
\end{equation}
Fig. 9 (c) presents the evolutional behaviors of \emph{aveQ} along the blue line in Fig. 1 for $N=5$, $N=10$ and $N=25$. The values of \emph{aveQ} are somewhat robust to boson number $N$. For the prolate shape, it is around -1.2 while for the oblate shape, it is around 0.6. The asymmetry of the prolate-oblate shape transition can be seen clearly via this order parameter. The average deformation of the prolate shape is nearly twice the one of the oblate shape. This is an important result in this paper.

\begin{figure}[tbh]
\includegraphics[scale=0.27]{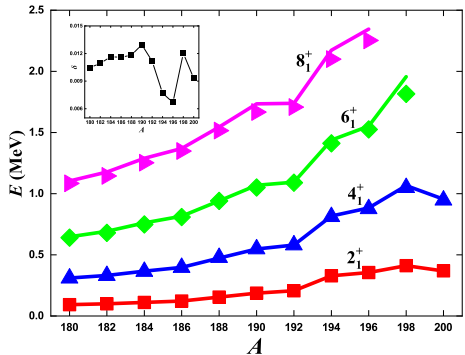}
\caption{Experimental excitation energies (symbol) and theoretical results obtained from the SU3-IBM calculations (line) for the yrast band in the Hf-Hg region. The inset is the coefficients $\delta$.}
\end{figure}

The dimensionless order parameter $\zeta$ is introduced to reveal the emergence of the new $\gamma$-softness, which is
\begin{equation}
\zeta=Q^{2}_{2_{1}^{+}}/B(E2;2_{1}^{+}\rightarrow 0_{1}^{+}).
\end{equation}
This quantity is not relevant to the effective charge $q$. Fig. 10 presents the evolutional behaviors of $\zeta$ along the blue line in Fig. 1 for $N=5$, $N=10$ and $N=25$. In the prolate side, this quantity is around 4.2 and robust to $N$. In the oblate side, it changes from 3.6 to 4.1 when $N$ increases. In the transitional region, it can be zero for the $\gamma$-softness.
Thus the emergence of the $\gamma$-softness is obvious for $\zeta$.

Another order parameter $\beta$ is also used to describe the qradrupole deformation \cite{Wang08,Tikkanen01,Swiatecki95}, which is defined as
\begin{equation}
\beta=\frac{4\pi}{3ZR_{0}^{2}}\left[\frac{B(E2;0_{1}^{+}\rightarrow 2_{1}^{+})}{e^{2}}  \right]^{1/2},
\end{equation}
where $Z$ is the proton number, $R_{0}$ is the mean radius of nucleus, and $e$ is the charge. Fig. 11 presents the evolutional behaviors of $\beta$ along the blue line in Fig. 1 for $N=5$, $N=10$ and $N=25$. It can be seen that, for definite $N$, the value of the $\beta$ at the prolate side is nearly twice than the ones at the oblate side, which means that the prolate-oblate shape transition is not symmetric and unlike previous IBM in the inset. This evolutional trend is in accordance with the results obtained in the energy density functional theories \cite{Nomura11t,Nomura11,Li21}.

All the above quantities are chosen to study the prolate-oblate shape phase transitions because they can be compared with existing experimental data in the Hf-Hg region.

\section{Prolate-oblate shape transition in the Hf-Hg region}

Prolate-oblate shape transition in the Hf-Hg region has been discussed in Ref. \cite{jolie03,Wang08,Nomura11t,Nomura11,Zhang12,Li21}. Earlier discussions were based on the $SU(3)-O(6)-\overline{SU(3)}$ evolution path in previous IBM, which is a mirror symmetric description for the prolate shape and the oblate shape \cite{jolie03,Wang08}. The results with energy density functional revealed that, this evolutional behaviors from the prolate shape to the oblate shape is in fact an asymmetric one \cite{Nomura11t,Nomura12,Li21}. Ref. \cite{Zhang12} first provided an asymmetric description within the $SU(3)$ limit, which is a degraded one. Ref. \cite{Fortunato11} showed that, an effective asymmetric description for the prolate-oblate shape phase transition can be realized in the cubic-$Q$ Hamiltonian in the large $N$ limit. In this paper, concrete numerical calculations are performed in the SU3-IBM, and these results further confirm the asymmetric behaviors. This means that the new model can offer a better description of the properties of realistic nuclei.

\begin{table}[tbh]
\caption{\label{table:expee}  Parameters $\kappa$, $\delta$ and effect charge $q$ used to fit the experimental data from $^{180}$Hf to $^{200}$Hg.}
\begin{tabular}{ccccc}
\hline
\hline
             $ ~~   $ & $\kappa$               \ &$\delta$ (MeV)        \ & $q$ ($\sqrt{W.u.}$)     \\
 \hline
   \\
$   ^{180}$Hf$~~   $ & 0.717               \ & 0.01047           \ & 1.413            \\
$   ^{182}$W$ ~~   $ & 0.750               \ & 0.01093           \ & 1.433           \\
$   ^{184}$W$ ~~   $ & 0.807               \ & 0.01165           \ & 1.475            \\
$   ^{186}$W$ ~~   $ & 0.873               \ & 0.01159           \ & 1.596             \\
$   ^{188}$Os$~~   $ & 0.972               \ & 0.01183           \ & 1.551             \\
$   ^{190}$Os$~~   $ & 0.987               \ & 0.01296           \ & 1.691             \\
$   ^{192}$Os$~~   $ & 0.993               \ & 0.01116           \ & 1.855         \\
$   ^{194}$Pt$~~   $ & 1.335               \ & 0.00769           \ & 2.249                \\
$   ^{196}$Pt$~~   $ & 1.404               \ & 0.00668           \ & 2.385                 \\
$   ^{198}$Hg$~~   $ & 1.707               \ & 0.01203           \ & 2.494               \\
$   ^{200}$Hg$~~   $ & 2.307               \ & 0.00931           \ & 2.926              \\
\\
\hline
\hline
\end{tabular}
\end{table}

 When fitting the nuclei from $^{180}$Hf to $^{200}$Hg, the total fitting parameter \emph{c} is set to 1 for clarity, $\eta=0.5$, thus the adjustable parameter in Hamiltonian (1) is the $\kappa$. A detailed fitting will be done in future for other $SU(3)$ higher-order interactions are introduced. In order to better compare with the energies within the yrast band, the angular momentum interaction $\delta \hat{L}^{2}$ should be introduced into the Hamiltonian (1), where $\delta$ is the second parameter. The two parameters $\kappa$ and $\delta$ can be determined by the energies of the $2_{1}^{+}$ and $4_{1}^{+}$ states for one specific nucleus, then the effective charge $q$ can be also determined for determining the reduced transitional probability $B(E2;2_{1}^{+}\rightarrow 0_{1}^{+})$. Table I presents the three parameters $\kappa$, $\delta$ and effect charge $q$ used to fit the experimental data.

\begin{figure}[tbh]
\includegraphics[scale=0.27]{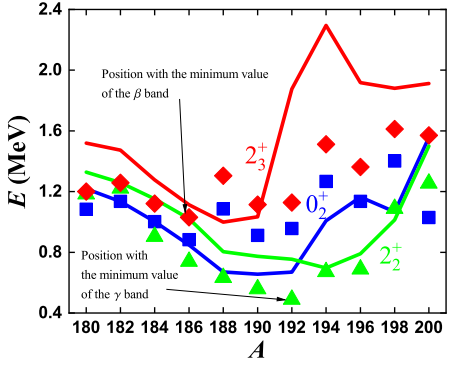}
\caption{Experimental excitation energies (symbol) and theoretical results obtained from the SU3-IBM calculations (line) for the $0_{2}^{+}$, $2_{2}^{+}$ and $2_{3}^{+}$ states in the Hf-Hg region. }
\end{figure}

Fig. 12 compares the theoretical values with experimental level energies in the yrast band in the Hf-Hg region. For two parameters $\kappa$ and $\delta$ are used, the $2_{1}^{+}$ and $4_{1}^{+}$ states are surely fitted well (errors come from numerical calculations), but the $6_{1}^{+}$ and $8_{1}^{+}$ states are also fitted well. When mass number $A$ increases, the shape parameter $\kappa$ also becomes larger (see Table I). Thus our new theory can provide a self-contained description for the prolate-oblate shape phase transition from $^{180}$Hf to $^{200}$Hg. For $^{180}$Hf, $^{182-186}$W and $^{188-192}$Os, their parameter $\kappa < 1.314$, so their shapes are prolate. For $^{194, 196}$Pt and $^{198, 200}$Hg, the values $\kappa > 1.314$, so they are oblate shapes. For $^{194}$Pt, its $\kappa$ is 1.335, very close to the critical point 1.314, so $^{194}$Pt is the critical nucleus for the prolate-oblate shape phase transition, which was also pointed out in \cite{jolie03,Bonatsos04}. For $^{196}$Pt, its $\kappa$ value 1.404 is just the location of the $B$ point, which is discussed in \cite{Wang23}.

Fig. 13 compares the theoretical values with experimental energies of the $0_{2}^{+}$, $2_{2}^{+}$ and $2_{3}^{+}$ states. The overall evolutional behavior can be reproduced by the new model qualitatively.  $2_{2}^{+}$ state is the bandhead of the $\gamma$ excitation band, and it is a key indicator for the prolate-oblate shape phase transition. It is shown that, the theoretical values can well produce the evolutional trends of the realistic nuclei except some values are larger. $0_{2}^{+}$ is the bandhead of the $\beta$ excitation band, and the deviations between the theoretical values and the experimental data should be further discussed. The position of the minimum value of the $2_{2}^{+}$ state is at $^{192}$Os, and the position of the minimum value of the $2_{3}^{+}$ state is at $^{186}$Pt. Theoretically, the two positions are $^{194}$Pt and $^{188}$Os. Thus the new model indeed reproduce the realistic evolutional behavior, but it is slightly insufficient quantitatively. Introducing more $SU(3)$ higher-order interactions may improve the fitting effect, which will be discussed in future.
Even with the deficiencies, the fitting results clearly show that the idea of the new model is correct, the prolate-oblate shape phase transtion is indeed asymmetric, and the description of previous IBM is inappropriate.

In Fig. 14 , the order parameter $E_{4/2}$ shows that realistic prolate-oblate shape phase transition is an asymmetric one, and it is completely consistent with our theory. At the prolate side, $E_{4/2}$ is nearly 3.3, while at the oblate side, it is nearly 2.6. At the critical point, the value is nearly 2.5 (see previous Fig. 4). It should be noticed that, previous IBM-1 can not offer such asymmetric behaviors. For $^{198, 200}$Hg, they are really oblate shapes. The order parameter $E_{3/2}$ is qualitatively reproduced. In Fig. 15 the experimental data of $E_{4/2}$ and $E_{3/2}$ are shown for the oblate nuclei with positive quadrupole moment $Q_{2_{1}^{+}}$. $E_{4/2}$ values are around 2.5, and $E_{3/2}$ values rapidly increases from 1.2 at $^{200}$Hg to 2.0 at $^{196}$Pt. These features are consistent with the new model.

Above all, the fitted results with two parameters really reproduce the level features from $^{180}$Hf to $^{200}$Hg at a better level, which can not be done in previous studies.

\begin{figure}[tbh]
\includegraphics[scale=0.27]{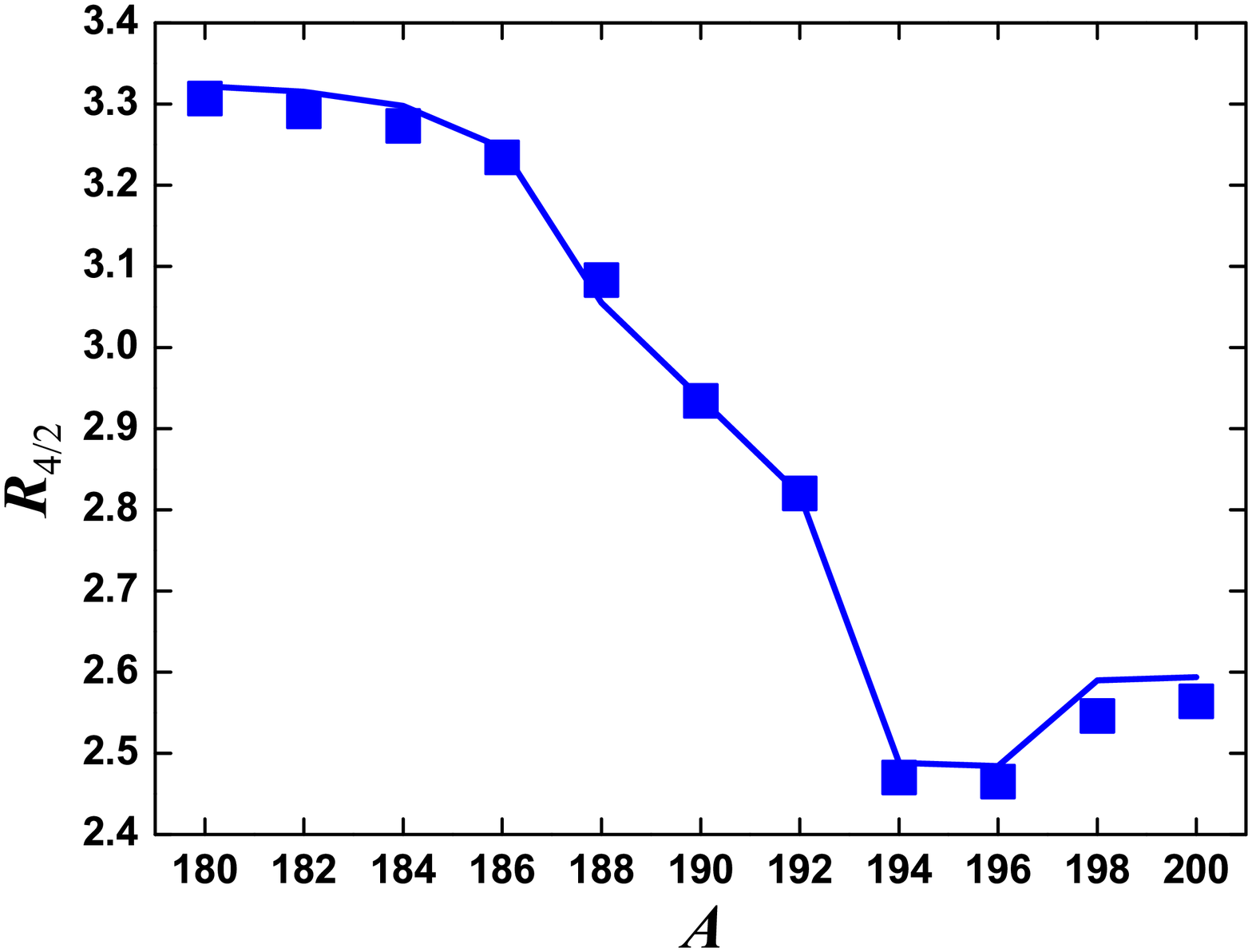}
\includegraphics[scale=0.27]{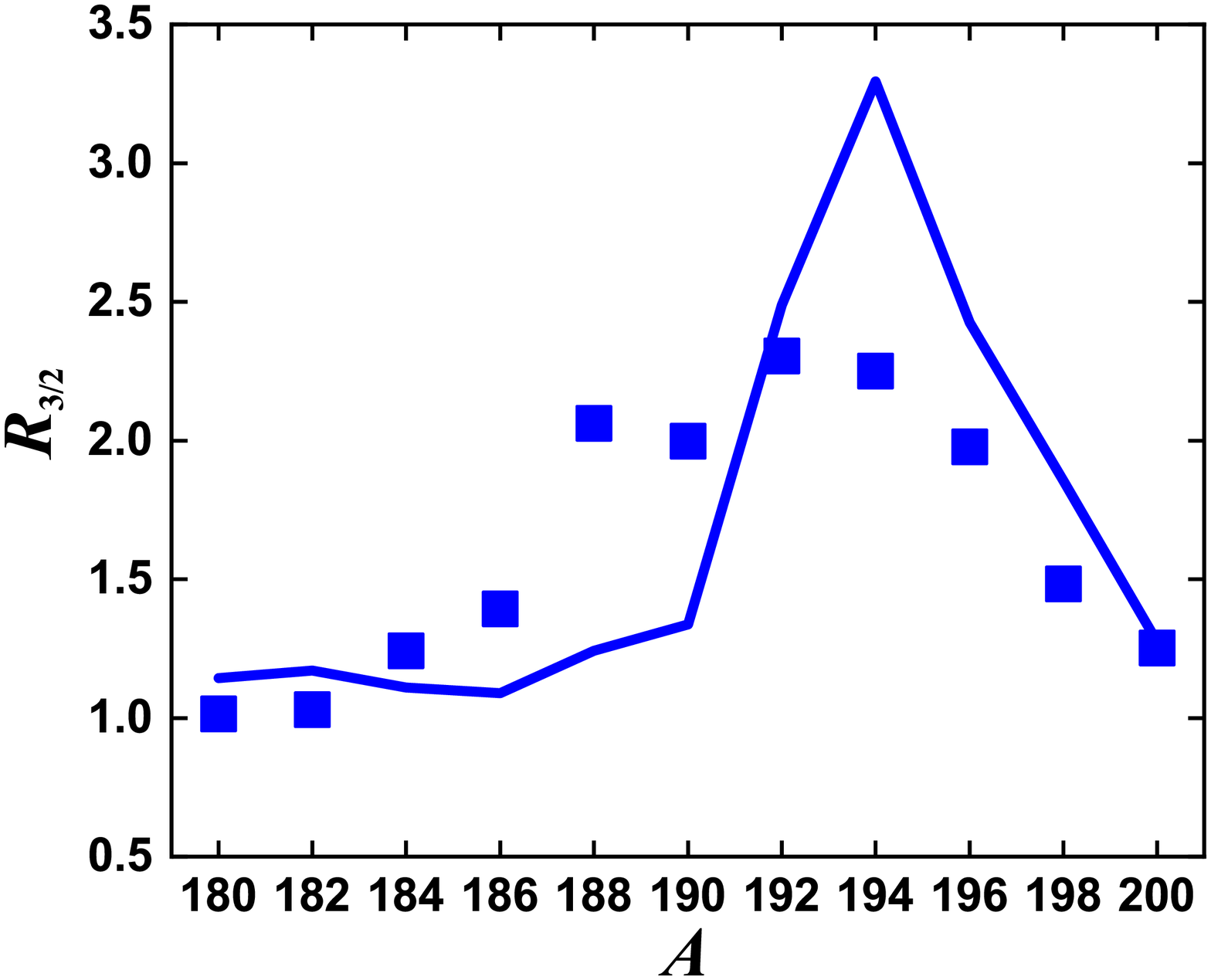}
\caption{Experimental data (symbol) and theoretical results obtained from the SU3-IBM calculations (line) for $R_{4/2}$ and $R_{3/2}$ in the Hf-Hg region. }
\end{figure}

\begin{figure}[tbh]
\includegraphics[scale=0.27]{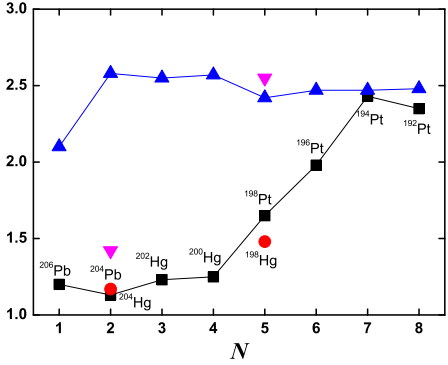}
\caption{Experimental data for $R_{4/2}$ (blue and magenta) and $R_{3/2}$ (black and red) of the oblate nuclei in the Pt-Pb region as the function of boson number $N$.  }
\end{figure}

\begin{figure}[tbh]
\includegraphics[scale=0.27]{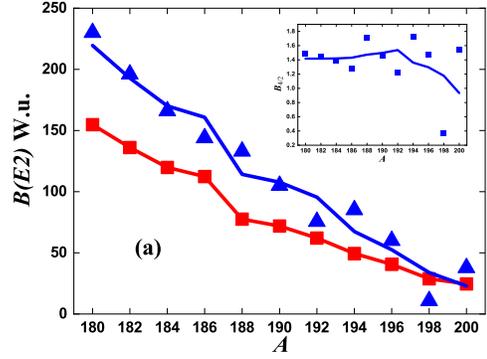}
\includegraphics[scale=0.27]{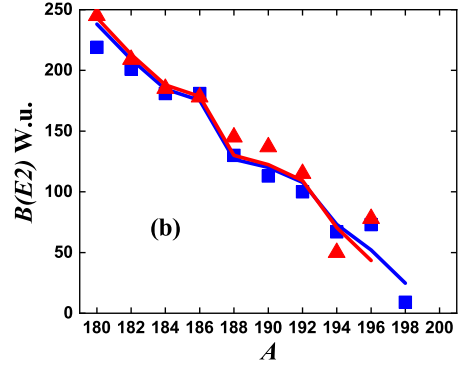}
\caption{(a) Experimental data (symbol) and theoretical results obtained from the SU3-IBM calculations (line) for the $B(E2;2_{1}^{+}\rightarrow 0_{1}^{+})$, $B(E2;4_{1}^{+}\rightarrow 2_{1}^{+})$ in the Hf-Hg region. The inset in (a) presents the  experimental data (symbol) and theoretical results obtained from the SU(3)-IBM calculations (line) for $B_{4/2}$. (b) Experimental data (symbol) and theoretical results obtained from the SU3-IBM calculations (line) for the $B(E2;6_{1}^{+}\rightarrow 4_{1}^{+})$ and $B(E2;8_{1}^{+}\rightarrow 6_{1}^{+})$.}
\end{figure}

Fig. 16 (a)-(c) present the reduced transitional probabilities $B(E2;2_{1}^{+}\rightarrow 0_{1}^{+})$, $B(E2;4_{1}^{+}\rightarrow 2_{1}^{+})$, $B(E2;6_{1}^{+}\rightarrow 4_{1}^{+})$ and $B(E2;8_{1}^{+}\rightarrow 6_{1}^{+})$ within the yrast band from $^{180}$Hf to $^{200}$Hg. The overall evolutional trends can be reproduced well. $^{198}$Hg shows a behavior like $B(E2)$ anomaly, which needs further investigations. For $N$ is small, only some low-lying levels can be discussed. The inset in Fig. 16 (a) shows the evolutional behaviors of the order parameter $B_{4/2}$, which is not sensitive to the prolate-oblate shape phase transition.

A critical order parameter for the prolate-oblate shape phase transition is the reduced transitional probabilities $B(E2;2_{2}^{+}\rightarrow 2_{1}^{+})$. This quantity is small for the prolate shape or the oblate shape, while it is large for the $\gamma$-soft critical region. Fig. 17 presents a distinct phase transition behavior. The overall evolutional behaviors can be clearly reproduced, but a systematic deviation emerges. Comparing the curve of the experimental data, the theoretical fitting curve are shifted to the right a bit. This suggests that one $SU(3)$ higher-order interaction should be introduced to further improve the accuracy of fitting.

\begin{figure}[tbh]
\includegraphics[scale=0.27]{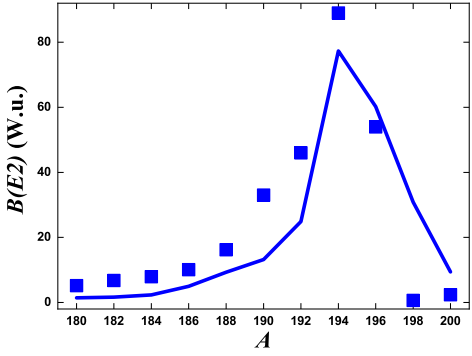}
\caption{Experimental data (symbol) and theoretical results obtained from the SU3-IBM calculations (line) for the $B(E2;2_{2}^{+}\rightarrow 2_{1}^{+})$ in the Hf-Hg region. }
\end{figure}

Fig. 18 (a)-(c) present the evolutional behaviors of the quadrupole moment $Q_{2_{1}^{+}}$ and $Q_{2_{2}^{+}}$, the two order parameter $aveQ$ and $\zeta$. These quantities are good indicators for the prolate-oblate shape phase transition. In Fig. 18 (a), from the sign of the values of the quadrupole meoment $Q_{2_{1}^{+}}$, it is negative for $^{180}$Hf, $^{182-186}$W, $^{188-192}$Os, and positive for $^{194, 196}$Pt, $^{198, 200}$Hg, thus a prolate-oblate shape phase transition really occurs. The evolutional behavior of $Q_{2_{2}^{+}}$ can be reproduced qualitatively. The experimental value of $^{184}$W is nearly zero, which can not be explained in previous IBM. In the new model, this feature can occur at $^{186}$W. For revealing the asymmetry of the two shapes, In Fig. 18 (b), the values of $aveQ$ at the prolate side is nearly 1.2, while it is nearly 0.6 at the oblate side. The theoretical results and the experimental data agree at a high level. The small theoretical values in $^{194, 196}$Pt may need other $SU(3)$ higher-order interactions, such as discussed in \cite{Pan18,Li22}. The theoretical values of the order parameter $\zeta$ can reproduce the overall evolutional features in Fig. 18 (c), and the emergence of the new $\gamma$-softness is prominent. The asymmetry of the prolate-oblate shape transition can be also seen for the order parameter $\beta$ in Fig. 19.

\begin{figure}[tbh]
\includegraphics[scale=0.27]{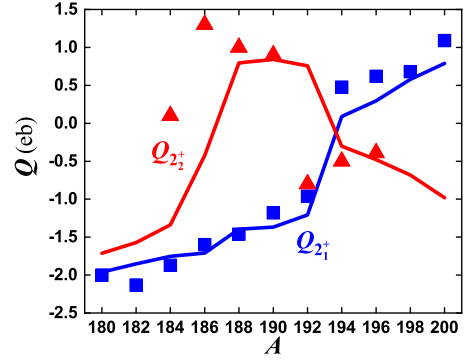}
\includegraphics[scale=0.27]{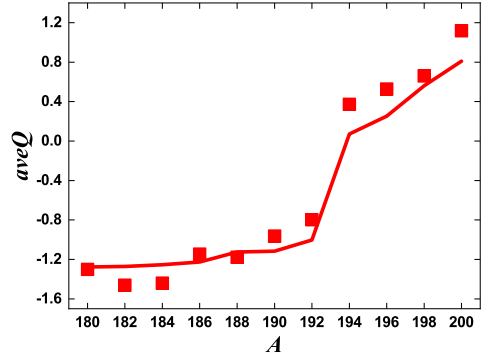}
\includegraphics[scale=0.27]{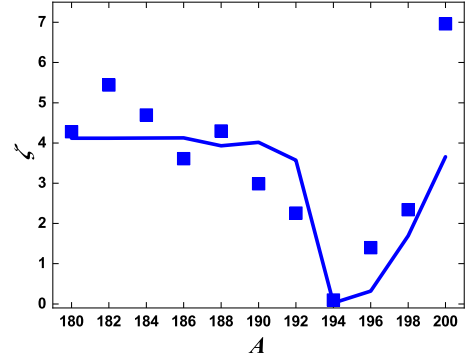}
\caption{Experimental data (symbol) and theoretical results obtained from the SU3-IBM calculations (line) for the quadrupole moments, $aveQ$ and $\zeta$ in the Hf-Hg region.  }
\end{figure}

\begin{figure}[tbh]
\includegraphics[scale=0.27]{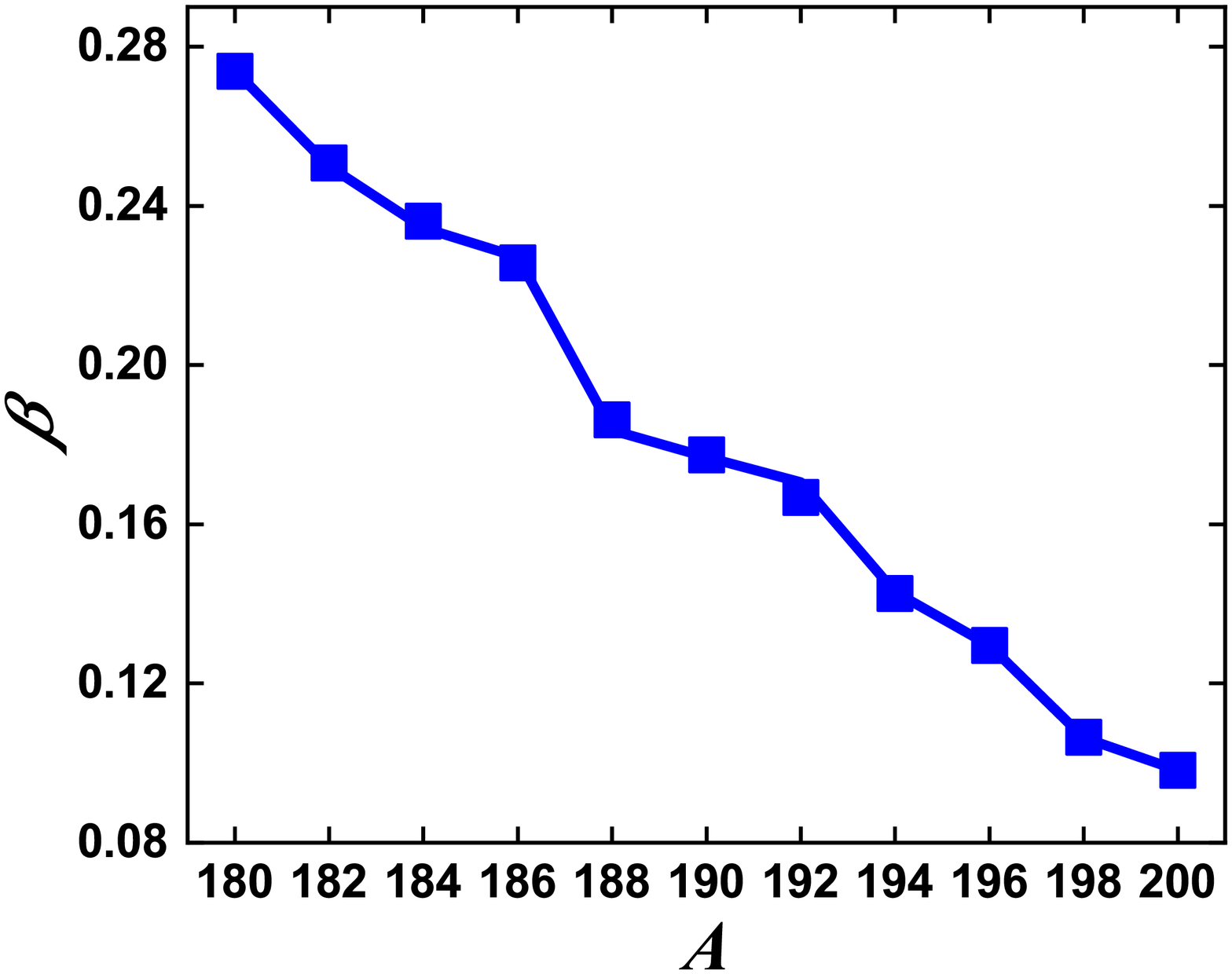}
\caption{Experimental data (symbol) and theoretical results obtained from the SU3-IBM calculations (line) for $\beta$ in the Hf-Hg region.  }
\end{figure}

\section{Some discussions about the prolate-oblate shape phase transition and the SU3-IBM}

$SU(3)$ higher-order interactions play a critical role in the new $SU(3)$-IBM. In previous studies on IBM-1, higher-order interactions are found to be necessary for describing rigid triaxial deformation of the ground state of a nucleus \cite{Isacker81,Isacker84}, that 6-d interaction  $[d^{\dag}d^{\dag}d^{\dag}]^{(L)}\cdot[\tilde{d}\tilde{d}\tilde{d}]^{(L)}$ can induce triaxiality. Then in the $SU(3)$ limit, the higher-order $SU(3)$ conserving interactions were investigated to remove the degeneracy of the $\gamma$ band and $\beta$ band \cite{Isacker85}. The degeneracy is a basic feature for the $SU(3)$ Casimir operators. Up to second-order interaction, only $-\hat{C}_{2}[SU(3)]$ and $\hat{L}^{2}$ are adopted. In Ref. \cite{Isacker85}, these higher-order interactions were only used as perturbations, and the third-order Casimir operators $\hat{C}_{3}[SU(3)]$ were not considered seriously. A key step with deeper physical meanings was an algebraic realization of the rigid asymmetric rotor within the $SU(3)$ limit of the IBM \cite{Isacker00, zhang14}. Recently, this realization was used to describe the $B(E2)$ anomaly \cite{Zhang22}. $SU(3)$ higher-order interactions were also investigated in Ref. \cite{Rowe77,Draayer85,Leschber88,Elliott86,Kota20}. These works laid the foundation for a comprehensive understanding of the $SU(3)$ limit of the IBM, but for a long time they did not attract enough attentions. In partial dynamical symmetry \cite{Leviatan11}, one symmetry is obeyed by some states and is strongly broken in others. Higher-order terms can induce such effects \cite{Leviatan09,leviatan13,Leviatan16,Leviatan17,Leviatan18,Leviatan21}. Third-order interaction $(\hat{Q}_{0}\times\hat{Q}_{0}\times\hat{Q}_{0})^{(0)}$ can show a rotational spectrum \cite{Isacker99}, where $\hat{Q}_{0}$ is the quadrupole operator in the $O(6)$ limit, which was further studied by \cite{Rowe05, Dai12}. Especially in the new developments \cite{Wang22,Wang20,Zhang22, Fortunato11,Zhang12,zhang14}, $SU(3)$ higher-order interactions can be used to describe some realistic anomalies in nuclear structure, and provide a new description for the oblate shape and the rigid triaxial shape, thus these interactions are of practical significance. Even in the IBM-2, higher-order interaction  $[d^{\dag}d^{\dag}d^{\dag}]^{(L)}\cdot[\tilde{d}\tilde{d}\tilde{d}]^{(L)}$ plays an important role for understanding the $\gamma$ excited band \cite{Nomura12}.

SU3-IBM can provide a preferred theoretical framework for simultaneously investigating the oblate shape, new $\gamma$-softness experimentally found, $B(E2)$ anomaly and the prolate-oblate shape phase transition. In this model, various rigid quadrupole deformations can be described within the $SU(3)$ limit using a unified way. When the $\hat{n}_{d}$ interaction is added, various $\gamma$-softness can occur, even having partial $O(5)$ partial dynamical symmetry, which is the most important finding in the new theory. Such a self-consistent description, albeit introducing higher-order interactions, is mathematically more concise and can better describe the properties of the actual nuclei. In previous IBM-1, the $O(6)$ limit describes the $\gamma$-related rotational mode, so it has higher symmetry than the $SU(3)$ limit. Recent experimental studies on the $O(6)$ symmetry have found that, in actual nuclei this symmetry is broken \cite{Werner10,Werner11,Morrison20,Kisyov22,Clement23}. If the forth-order $\hat{C}_{2}^{2}[SU(3)]$ interaction is added in the SU3-IBM, new transitional behaviors like the $U(5)$-$O(6)$ shape phase transition in the IBM-1 can be obtained \cite{Zhou23}, and $E(5)$-like or $O(6)$-like spectra can emerge in the new model. These should be further investigated in future. New $\gamma$-softness in the SU3-IMB has lower symmetry, which seems a more realistic description.

From these new studies, it can be seen that how to describe the oblate shape is the key point to the new understanding. The perspective that triaxiality results from the competition between the prolate shape and the oblate shape can deepen our understanding on the nuclear structure. The new theory is precisely inspired by this idea \cite{Wang22}. In previous IBM-1, the oblate shape is described by the $\overline{SU(3)}$ symmetry, which can be seen as a mirror image of the $SU(3)$ limit, thus the critical point $O(6)$ symmetry can be exactly unrelated to the $\gamma$ degree of freedom. Thus in the old description, there is an exact prolate-oblate mirror symmetry. Although it looks better from a mathematical perspective, the actual evolution of nuclear structure does not support this view. The emergence of new $\gamma$-softness can provide us a new way to reconsider the realistic prolate-oblate shape phase transition. The prolate-oblate mirror symmetry does not exist, as shown in this paper. In the various quadrupole deformations, the oblate shape is the least studied. In most studies of nuclear structure, the oblate shape is usually regarded as an image of the prolate shape. In the new series of studies, the property of the oblate shape is placed in a remarkable position, which is no longer in the shadow of the prolate shape and may have different performance. In this way, the oblate shape and the prolate shape have different formation mechanisms. Thus the realistic $\gamma$-softness is not the one described in the $O(6)$ limit. More researches on the oblate nuclei are required, both experimentally and theoretically. New works show that, when the proton number or neutron number approaches the magic number, the nuclei can have an oblate shape, rather than a spherical shape.

Spherical nucleus puzzle and $B(E2)$ anomaly exacerbate the shift of this new idea. These odd experimental results conflict with the old ideas based on spherical vibrations, and are difficult to interpret by previous nuclear structure theories. In Ref. \cite{Heyde16}, Heyde and Wood said:  "The emerging picture of nuclear shapes is that quadrupole deformation is fundamental to achieving a unified view of nuclear structure. While it has now long been recognized that many nuclei are deformed, the reference frame for nuclear structure discussion has been spherical shapes. We would argue that a shift in perspective is needed: \emph{sphericity is a special case of deformation}. Thus, we argue that the reference frame must be fundamentally one of a deformed many-body system. The dominance of spherical shapes in formulating descriptions of nuclear structure has been dictated by the preferred basis used in constructing many-particle wave functions." The new theory supports the idea in a very consistent way. This may seem very surprising, because there is not much evidence to support the emergence of the new theory. Ref. \cite{Wang22} can be seen as an extended investigation of Ref. \cite{Fortunato11}, but no one realizes this new $\gamma$-softness until the implied results comes out. The most important result is the connections between the new $\gamma$-softness and the normal states of $^{110}$Cd. $\gamma$-soft behaviors revealed by experimental data \cite{Garrett12,Batchelder12,Garrett19,Garrett20} seem to be just the new $\gamma$-softness found in the SU3-IBM. Moreover, the new theory is still the only one that can explain the $B(E2)$ anomaly \cite{Wang20,Zhang22}. For $^{110-114}$Cd are three nuclei with spherical nucleus puzzle and $^{114}$Te, $^{114}$Xe are two nuclei with $B(E2)$ anomaly, a deep connection between the two anomalies must exist. $^{166-172}$Os and $^{162-168}$W have similar evolutional behaviors from $B(E2)$ anomaly to normal case, which are completely different from the existing concept of nuclear structure evolution. Experimental researches on $^{162,164}$Os and $^{158,160}$W may reveal more relationships between the spherical nucleus puzzle and the $B(E2)$ anomaly.

In the present SU3-IBM, the proton pair and the neutron pair are not distinguished like previous IBM-1. Thus a direct important extension is to construct a theory in which the proton pair and the neutron pair are different to treat. This extended theory can be called SU3-IBM-2 like IBM-2, and the present model is tabbed as SU3-IBM-1. How to deal with these $SU(3)$ higher-order interactions in the SU3-IBM-2 is the key step. One simple way is to view these $SU(3)$ higher-order interactions as merely the ones belonging to the proton pair or the neutron pair, and the two-body interaction between the proton pair and the neutron pair is the usually used on in the previous IBM-2.
This idea requires further numerical implementation, and it should explain the new $\gamma$-softness and the $B(E2)$ anomaly at a better level.

Realizing the new $\gamma$-softness in the geometric model will be an important step for extending the new idea \cite{Fortunato05,Fortunato16}. A numerical calculation of the geometric model with the $\gamma$-soft potential in Ref. \cite{Fortunato11} may give the new $\gamma$-soft behaviors. This result can be further used to resolve the $B(E2)$ anomaly in the geometric model like \cite{Oulne22}. This will also can help us to understand the oblate shape to correct the finite $N$ effect in the SU3-IBM.

Energy density functional theory can provide the microscopic description of the various nuclear structure evolutions \cite{Ring11,Ring19}. Recently a novel way of determining the parameters of the Hamiltonian of previous IBM based on the mean-field models was given in \cite{Nomura08,Nomura10}, which offers a microscopic foundation of the IBM and can provides a deep insight on the nuclear structure evolutions \cite{Nomura11t,Nomura11,Nomura12,Nomura21,Nomura22}. Similar ideas can be also used in the new model, and the part parameters of the Hamiltonian in the SU3-IBM can be obtained microscopically. This novel way can especially facilitate the understanding of the new theory.

Some new ideas also emerges recently. The shell model has some deficiencies \cite{Wood22}, the role of the single-particle energy gap has bee also stressed \cite{Otsuka19,Otsuka22}, and oblate-prolate phase mechanism in the Te-Ba region is also studied \cite{Shimizu23}. We expect our new model can provide more insights on these new ideas.

\section{Conclusions}

Prolate-oblate shape phase transition in the Hf-Hg region is studied by the interacting boson model with $SU(3)$ higher-order interactions (SU3-IBM). This is the first par of this work. This work can be seen as a further numerical investigations of the theoretical work in \cite{Fortunato11}, in which the large $N$ case was first studied. In previous IBM with $\overline{SU(3)}$ symmetry as the description of the oblate shape, there exists a mirror symmetry between the prolate shape and the oblate shape from the spectra perspective. However, this feature can not be found in realistic nuclei. In our paper, the calculation results in the new model support realistic evolutional behaviors. This greatly changes our understanding of the algebraic model.

Further works will be done in future for an improved understanding of the prolate-oblate shape phase transition and the new $\gamma$-softness. Firstly, an overall study on the oblate nuclei in the Pt-Pb region is needed. Secondly, the evolutional behaviors at the prolate side in the W-Os isotopes are also vital for us to understand this shape transition. Further studies need to introduce more $SU(3)$ higher-order interactions,  and there may be a lot of details that needs to be discussed clearly.

In conclusion, our work supports the rationality of the new model. Such work will likely revolutionize our understanding of nuclear structure.

\section{ACKNOWLEDGMENT}

 This research is supported by Science and Technology Research Planning Project of Education Department of Jilin Province (JJKH20210526KJ).

\end{document}